\documentclass[reprint, superscriptaddress]{revtex4-1}
\usepackage[utf8]{inputenc}
\usepackage[T1]{fontenc}
\usepackage{graphicx}
\usepackage{grffile}
\usepackage{longtable}
\usepackage{wrapfig}
\usepackage{rotating}
\usepackage[normalem]{ulem}
\usepackage{amsmath}
\usepackage{textcomp}
\usepackage{amssymb}
\usepackage{capt-of}
\usepackage{hyperref}
\usepackage{booktabs}
\usepackage{float}
\usepackage{times}
\usepackage{lipsum}
\usepackage{float}
\usepackage{setspace}
\usepackage{blindtext}
\usepackage{tikz}
\usepackage{hyperref}
\date{}
\title{}
\hypersetup{
 pdfauthor={},
 pdftitle={},
 pdfkeywords={},
 pdfsubject={},
 pdfcreator={Emacs 27.1 (Org mode 9.3.7)}, 
 pdflang={English}}
\begin{document}

\newcommand{\cmt}[1]{{\null}}
\newcommand{\ts}[1]{\textsuperscript{#1}}

%% \title{Self-consistent s-p-d tight-binding model of Fe, parameterised against results based on QSGW and DFT approximations}
\title{Development of data-driven \textit{spd} tight-binding models of Fe -- parameterisation based on QSGW and DFT calculations including information about higher-order elastic constants.}
\author{Bartosz Barzdajn}
\affiliation{The University of Manchester, Department of Materials}
\author{Alexander M. Garrett}
\affiliation{The University of Manchester, Department of Materials}
\author{Thomas M. Whiting}
\affiliation{Imperial College, Department of Materials}
\author{Christopher P. Race}
\affiliation{The University of Manchester, Department of Materials} 

\date{\today}
\begin{abstract}
Quantum-mechanical (QM) simulations, thanks to their predictive power, can provide significant insights into the nature and dynamics of defects such as vacancies, dislocations and grain boundaries. These considerations are essential in the context of the development of reliable, inexpensive and environmentally friendly alloys. However, despite significant progress in computer performance, QM simulations of defects are still extremely time-consuming with ab-initio/non-parametric methods. The two-centre Slater-Koster (SK) tight-binding (TB) models can achieve significant computational efficiency and provide an interpretable picture of the electronic structure. In some cases, this makes TB a compelling alternative to models based on abstraction of the electronic structure, such as the embedded atom model. The biggest challenge in the implementation of the SK method is the estimation of the optimal and transferable parameters that are used to construct the Hamiltonian matrix. In this paper, we will present results of the development of a data-driven framework, following the classical approach of adjusting parameters in order to recreate properties that can be measured or estimated using ab-initio or non-parametric methods. Distinct features include incorporation of data from QSGW (quasi-particle self-consistent GW approximation) calculations, as well as consideration of higher-order elastic constants. Furthermore, we provide a description of the optimisation procedure, omitted in many publications, including the design stage. We also apply modern optimisation techniques that allow us to minimise constraints on the parameter space. In summary, this paper introduces some methodological improvements to the semi-empirical approach while addressing associated challenges and advantages.
\end{abstract}
\maketitle

\section{Introduction}
\label{sec:orgc613cc2}
\subsection{Motivation}
\label{sec:org79177ab}
\cmt{Check if ill-posed is the right term.} We will present the development of our tight-bonding parametrisation methodology in the context of its applications. Our long-term objective is to investigate diffusion mechanisms in low-alloy reactor pressure vessel (RPV) steels and to understand precipitates and nano-cluster formation. These types of defects are important in the context of life-time studies of irradiated structural materials. A good model for this class of problems has to be able to approximate well enough formation and migration energies of point defects (vacancy and self-interstitial) while being able to predict the interactions of solute atoms with these defects. While such investigations are usually conducted using density-functional theory (DFT), they require significant computational resources, especially if we would like to consider multiple defects or dislocations  and explore the whole range of possibilities \cmt{provide suitable inputs to Monte Carlo simulations?}. Classical empirical potentials are an excellent interpolator, but require large data-sets to optimise and have limited predictive value. Amongst quantum-mechanical models, the two-centre Slater-Koster (SK) tight-binding (TB) method (\cite{Slater1954}) provides one of the simplest one-particle bases (\cite{Andersen1984}), namely the linear-combination of atomic orbitals (LCAO), with well-established rules to construct the Hamiltonian for s, p and d valence elements. This results in significantly increased computational efficiency and, for this reason, this classical technique is of ongoing interest within the material science community. However, the biggest associated challenge is to find transferable ("universal") parameters for constructing Hamiltonian matrix elements. Here we revisit this problem as an automated, simple, and data-driven \cmt{(meaning flexible)} method for finding parameters which we believe would be an asset to the community. We will explore one of the potential approaches which is similar to, among others, the Naval Research Laboratory Tight-Binding methodology (\cite{Cohen1994}). In this approach hopping integrals are adjusted to recreate features of the electronic structure, while parameters of the repulsive pair-potential are optimised for mechanical properties.
\subsection{The context}
\label{sec:orga1383c4}
We now address the problem of model selection. There are two major reasons for presenting an alternative parametrisation for Fe, a material which has been studied in the past. First, we are interested in studying low-alloy steels in the future. Secondly, Fe, as a ferromagnetic material, serves as a good case study for development of our techniques. We are trying to achieve a balance between the  predictive value of the model, speed and also the simplicity of the fitting process. In other words, we would like to find a convenient and fairly automated method of parametrisation. As a ferromagnetic transition metal, Fe requires appropriate treatment of electronic correlations. The challenge associated with the treatment of this phenomenon is best illustrated by Mehl et. al. (\cite{Mehl1996}), where a methodology that worked well for non-magnetic metals, failed in case of Fe, resulting in some elastic constants being negative. However, these correlations are not as "troublesome" as in the case of highly correlated systems that involve e.g. f-valence electrons. Methods based on the most common implementations of the density functional theory (Khon-Sham DFT, \cite{Kohn1965}) are fairly successful with 3d transition metals. This suggests that self-consistent tight-binding methods should be able to handle these type of materials as there is a direct link between such variants of the TB method (more information will be provided in the following section) and DFT with LDA (local density approximation) exchange-correlation functional \cite{Paxton2008}. In both methods (TB and DFT), two-electron states, responsible for the description of electronic correlations, are projected into a one-electron basis at each step of a self-consistency cycle. On the other hand, it is also well known that appropriate description of the bond-lengths requires, in the case of Fe, a generalised gradient approximation (GGA) to the exchange-correlation functional. This requirement might indicate that the TB picture is incomplete and, in some sense, the lack of physics needs to be compensated with adjustments of the model parameters \cmt{Implicitly this means adjustment of the starting electron density that has to be close-enough? How to word that?}. In our research we decided to use the polarisable-ion self-consistent tight-binding model \cite{Paxton2008}. As it that goes beyond a  simple Stoner model (\cite{Stoner1938}), \cmt{haven't read but found the eq. there, something that was mentioned by Paxton} we hope to achieve a good representation of magnetism and associated effects. Now we are left with two important choices: the basis and method of estimating transferable parameters. Examples of successful models of Fe include: a non-orthogonal \(spd\) basis by Bacalis et al. \cite{Bacalis2001} (no self-consistency), an orthogonal d-band by Liu et al. \cite{Liu2005}, a self-consistent non-orthogonal \(spd\) model by Paxton et al. \cite{Paxton2008}, another similar model with s-d basis by Paxton et. al. \cite{Paxton2010}, optimised orthogonal d-band by Madsen et. al.\cite{Madsen2011}, and self-consistent DFT-based orthogonal d-band model by Hatcher et. al. \cite{Hatcher2012}. \cmt{(that covers most of the people that potentially will review this:))} DFT-based parametrisation of the SK-TB has a great advantage as it addresses the main issue of finding suitable parameters (see also Horsefield et al. \cite{Horsfield1998} for more information on DFT-TB). However, DFT itself does not necessarily provide an optimal approximation and its quality may worsen when LCAO basis is used. \cmt{(not sure about this, I've made it up)} On the other hand, this can be compensated by refitting. Hence, we decided to focus our efforts on the semi-empirical scheme, as it can be used for refinement or as a stand-alone technique. An additional practical advantage is that the developed framework can then be used with any atomistic software package that implements the TB scheme. To minimise constraints we will be using a full \(spd\) basis, which will allow us also to recreate most of the features of the band structure.
\subsection{Methodology}
\label{sec:orgbe4050d}
As emphasised before, finding TB parameters when using a full \(spd\) non-orthogonal basis is a significant challenge. One of the key features of this paper is a presentation of the methodology used in the formulation of the optimisation problem. This includes the reasoning behind selecting fitness measures and objectives. For example, as an objective, we have used band structures estimated using three different methods: DFT with LDA, DFT with GGA and the quasi-particle self-consistent GW method -- QSGW (\cite{Kotani2007}). To be more specific, we have used band structures generated from the density updated in the QSGW self-consistency cycle \cite{gw_tutorial}.  We will demonstrate that it is possible to obtain a very good agreement with the QSGW results. This might be of particular interest for the community, as QSGW provides high-quality truly ab-initio results. We believe, that this approach may be used to improve existing parametrisations with results from a more complete, in the physical sense, models. Here, we are referring to the more accurate ab-initio treatment of electronic correlations. While the DFT implemented with the GGA (Generalised Gradient Approximation) works very well in case of Fe, the same approach may be adopted for materials for which it is not the case. Regarding our approach to selection of TB parameters, it wouldn't be possible to obtain such a good agreement with reference data without using somewhat novel approach, i.e. usage of multiple measures, each approaching different optimum, together with an efficient global optimisation algorithm. This way we reduced issues associated with ill-posed optimisation problems. We also show that by aiming to recreate third-order elastic constants we can sufficiently sample pair-wise interactions without applying large strains (large strains can move some atoms outside their interaction range). This is another way we reduce the degree to which the problem is ill-imposed as we exclude test-cases with a different number of considered neighbours. Also, we show that some deformation patterns can result in negative curvature in the energy-strain relationship even when other deformation patterns give reasonable elastic constants. By applying additional test-cases, necessary for the calculation of higher order elastic constants, we increase the robustness of the optimisation. Additionally, we demonstrate that using energy-strain curves estimated using the DFT method we can recreate experimental elastic constants. This approach has the advantage of being a much more computationally efficient way to quantify the goodness of fit. Furthermore, second derivatives are extremely sensitive to changes in model parameters and tend to make optimisation much more difficult. Finally, we were able to satisfy the objectives and obtain a reasonable transferability, while applying as little constraint as possible on parameters. However, the predictive value of our models, here represented by formation energies of point defects -- our main quantity of interest, may benefit from further improvement.

\section{Description of the model}
\label{sec:org396bc11}
The Tight-binding (TB) approximation to the many-body problem, which we consider in this paper, is a very well established technique. However, the name serves also as an umbrella term for many somewhat distinct methods. There are also many ways various effects can be included in models from the same category. For a general review we recommend publications by Paxton \cite{Paxton2009a}, Sutton and Balluffi \cite{sutton1997interfaces} or Finnis et al. \cite{Finnis2010}. The model we use is the self-consistent polarisable tight-binding model, as presented by Paxton et al. in \cite{Paxton2008} (see also Finnis et al. \cite{Finnis1997} and Sutton et al. \cite{Sutton1988}), as implemented in the Questaal package \cite{Pashov2019, PhysRevLett.81.5149}.
\cmt{For who this description is?} The theory behind the TB approximation we will address only briefly, in order to emphasise the key physical effects that are included and how this is done. The binding energy \(E_{\text{B}}\) (total energy minus energy of isolated atoms) of a system within this approximation, can be written as a sum \cite{sutton1997interfaces},
\begin{equation}
\label{eq:orgdb9beb0}
E_{\text{B}}=E_{\text{bond}} + E_{\text{prom}} + E_{\text{pair}},
\end{equation}
where \(E_{\text{bond}}\) is the covalent bond energy, \(E_\text{pair}\) represents Coulomb interactions between nuclei and \(E_{\text{prom}}\) represents changes in the energy due to differences in electronic configuration of a solid compared to isolated atoms. As argued by Sutton and Balluffi, \(E_{\text{pair}}\) may also encapsulate changes in electrostatic energy due to exchange-correlation interactions \cite{Sutton1988,sutton1997interfaces}.  Approach adopted e.g. by Paxton et al. differs in this context. The tight-binding Hamiltonian associated with the sum \(E_{\text{bond}} + E_{\text{prom}}\) is decomposed into two components
\begin{equation}
\hat{H}=\hat{H}_{1} + \hat{H}_{2},
\end{equation}
where \(\hat{H}_1\) represents non-interacting electrons in an effective potential and \(\hat{H}_2\) "\ldots{} describes electron-electron interactions and is constructed so as to represent second order terms in the expansion of the Hohenberg-Kohn density functional about a reference density \ldots{}" \cite{Paxton2008}. In this approach less is required of the pair-potential in reproducing the total energy. The model for electrons can be summarised more explicitly, yet in a fairly compact way, using a projection (\cmt{PHP? - something that removes overlap??; how overlap comes into this??}) to a Hubbard-like Hamiltonian \cite{Hubbard1963}. Using the formalism of the second quantisation and LCAO basis (e.g. \cite{Harrison1985} and \cite{Foulkes2016})
\begin{eqnarray}
\label{eq:org0efdf86}
\hat{H} & = & \sum_{\sigma}\sum_{\alpha\beta}h_{\alpha\beta\sigma}\hat{c}_{\alpha\sigma}^{\dagger}\hat{c}_{\beta\sigma}^{\,} \nonumber \\
 & + & \sum_{\sigma\sigma'}\sum_{\alpha\beta\gamma\delta}\mathcal{U}_{\alpha\beta\gamma\delta\sigma\sigma'}\hat{c}_{\alpha\sigma}^{\dagger}\hat{c}_{\beta\sigma'}^{\,}\hat{c}_{\gamma\sigma'}^{\dagger}\hat{c}_{\delta\sigma}^{\,},
\end{eqnarray}
where \(\hat{c}^{\dagger}\) and \(\hat{c}\) are creation and annihilation operators respectively. In the above it is assumed that all normalisation factors are included in the matrices. The Greek letters represent a combined index of site positions \(\vec{R}\) as well as quantum numbers \(l\) and \(m\), while \(\sigma\) indexes the spin. The first term in Equation  \ref{eq:org0efdf86}, represents elements of \(\hat{H}_1\)  i.e the classical non-self-consistent tight-binding approximation. The second term, corresponds to \(H_2\) and defines pair-wise Coulomb interactions. In the spirit of the Hubbard model, these are limited to intra-atomic interactions (see \cite{Harrison1985}) and the matrix \(\hat{\mathcal{U}}\) is non-zero only for diagonal and off-diagonal on-site elements. It consists of pair-wise terms, such as Coulomb integrals \cmt{$U$} (for combinations of a type  \(\alpha\alpha\beta\beta\)), and exchange \cmt{$J$} (\(\alpha\beta\beta\alpha\) type) only for atomic orbitals originating from the same atom. Inclusion of spin changes the span of \(\hat{H}\) to include the basis of Pauli matrices. In the given implementation, electron-electron contributions are included in a self-consistent manner, by contributing to all (diagonal and off-diagonal) on-site elements. Self-consistent interactions include Madelung and Hubbard potentials that yield appropriate shifts to the non-interacting Hamiltonian. The exchange is included via the Stoner parameter \cmt{$I$}. The cycle in the self-consistency loop continues until input and output charge distributions are self-consistent. The TB model can be regarded as a single-particle picture because within each iteration the model is projected into a one-particle basis. The basis set is defined in keeping with the Slater-Koster algorithm as are the rules for the creation of the initial Hamiltonian matrix \cite{Slater1954}. The core idea behind the Slater-Koster (SK) approximation is to represent Hamiltonian matrix elements (\(H_{1ij}\)) in terms of two-centre integrals. These integrals are given as
\begin{equation}
H_{1ij}=E_{\alpha\beta}=\int_{V} \phi_\alpha \left(\vec{r} - \vec{R}_I \right) \hat{H_1} \phi_\beta \left( \vec{r} - \vec{R}_J \right)\,\text{d}\vec{r},
\end{equation}
where \(\phi_{\alpha(\beta)}\) are atomic orbitals, \(\hat{H}_1\) is the Hamiltonian \(\hat{H}_1=\hat{T}+\hat{V}_\text{eff}\), consisting of the operator of kinetic energy of electrons \(\hat{T}\) and effective potential \(\hat{V_\text{eff}}\) -- approximated as a sum of spherically symmetric contributions centred at each atom of the unit cell. These integrals represent contributions to the bond from distinct pairs of atomic orbitals. In the Slater and Koster algorithm three-centre contributions are ignored. By setting the coordinate system for each pair in such a way that the \(z\) axis originates on one atom and connects with the other, it is possible to rewrite Hamiltonian matrix elements in terms of direction cosines and (fundamental) bond integrals, also referred to as hopping integrals.
For each atomic species, we need to find values of these integrals and their dependence on the bond length. In case of the self-consistent TB, the parameters define initial values of hopping integrals that are later updated during the cycle. The procedure continues until forces are self-consistent. Effects included in this procedure have been discussed in the previous paragraph. For more details we refer the reader to already cited works by Finnis et al. (\cite{Finnis1997}) and Paxton et al. \cite{Paxton2008}. As emphasised earlier, this method provides a significant computational efficiency while including a representation of the electronic correlations. \cmt{However, as will be discussed in the following section, finding optimal parameters poses a significant optimisation challenge.}

\subsection{The issue of parameters space dimensionality}
\label{sec:orgf3da5a2}

In the case of the \(spd\) basis, there are \(9\times9\) possible orbital pairs. In the two-centre approximation, due to the symmetry, this number reduces to 29 that need to be specified, which further can be represented using only 10 fundamental bond integrals (table 1. in Slater et al. \cite{Slater1954}). These integrals are designated as: \(ss\sigma\), \(sp\sigma\), \(pp\sigma\) \(pp\pi\), \(sd\sigma\), \(pd\sigma\), \(pd\pi\), \(dd\sigma\), \(dd\pi\) and \(dd\delta\). The first two Latin letters in this naming convention represent the type of orbitals that contribute to the bond, while the Greek letters correspond to the bond type. More information can be found in the original SK paper \cite{Slater1954}. 
Applying a non-orthogonal basis, with a single parameter controling the decay, doubles the degrees of freedom of the model. We also need to find diagonal on-site elements, one for each orbital type. The Hubbard-like \(U\) can be assumed constant (\(1\,\text{Ry}\) is a common value), although the Stoner parameter \cmt{$I$} needs to be optimised. For parameters associated with the electronic structure, we apply single-parameter exponential decay. In summary, with 10 orbitals in an spd basis, this gives us 40 parameters describing hopping and overlap as there are two parameters (magnitude and decay) that define a single quantity (overlap or hopping). Including the Stoner parameter and on-site energies results in 44 degrees of freedom (DOF) in total (with respect to the electronic structure). Additionally, we need to find parameters describing atomic pair-wise repulsion. This adds at least two extra DOF. Parameters can be found by comparing the results of TB calculations with experiments or ab-initio calculations. We are aiming to optimise a self-consistent model, based on the variational principle, on a non-orthogonal basis that trades off flexibility (completeness) for simplicity (LCAO rather than Slater determinant). This means that whatever relationship we aim to reproduce or whatever performance measure we apply, the optimisation problem is likely to be ill-posed i.e. it might be impossible to select the right constraints for parameters before optimisation is complete. In other words, depending on the objective and constraints, either there will be no optimal solution (over-constrained problem) or the solution will not be unique (under-constrained). Hence, one needs to carefully choose measures, objectives and stopping criteria.

\section{Defining fitness measures for band structure parameters}
\label{sec:org52b4441}

\begin{figure}[htb!]

\includegraphics[height=0.25\textwidth]{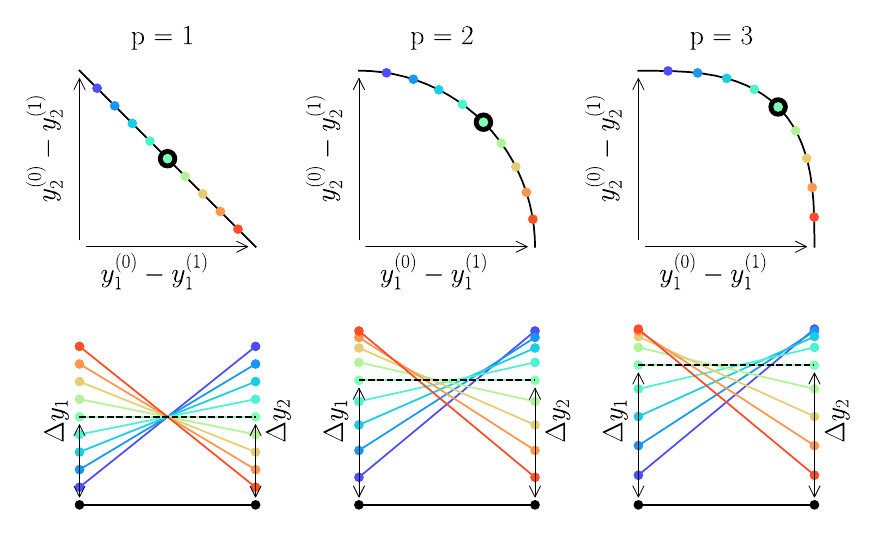}
\caption{\label{fig:org1d4418e}Distance between two vectors in two dimensions, on a unit sphere, that could also represent function values evaluated at two different points (bottom row). Here, \(\Delta y_i = y_i^{(1)} - y_i^{(0)}\). \cmt{objective as a fun. of p and p'?}}
\end{figure}

Band structures are inexpensive to calculate while providing a significant amount of information about the electronic structure. This is an important characteristic as it allows us to test different performance measures. Assuming that all measures correspond to the same optimum, at least with the right degree of constraint, it will be approached through a different path with the same algorithm. Given that initial constraints will most likely be sub-optimal, different measures will allow us to explore a wider range of candidate solutions and gain more confidance. We define the fitness function in a way analogous to the Minkowski distance i.e.
\begin{equation}
\label{eq:org13c10f9}
f_{\text{B}}\left(\vec{\alpha}_{\text{B}}\right) = \sum_{s\sigma}\left( A\left(s,\sigma\right)\sum_{nk}\left|{\boldsymbol \epsilon}_{nk}^{\left(\text{TB}\right)}\left(\vec{\alpha}_{\text{B}},s\right)-{\boldsymbol \epsilon}_{nk}^{\left(\text{ref.}\right)}\left(s\right)\right|^{p} \right)^{1/p'},
\end{equation}
where \(\vec{\alpha}_{\text{B}}\) is a tuple representing all parameters associated with the electronic structure, while \(\boldsymbol{\epsilon}^{(\text{TB)}}\left(\vec{\alpha}_{\text{B}},s\right)\) and \(\boldsymbol{\epsilon}^{(\text{ref.)}}\left(s\right)\) are matrices of TB and reference band structures respectively. The parameter \(s\) denotes both crystallographic structure and volumetric strain \(\gamma\). Possible values can be represented by a Cartesian product
\begin{equation}
\label{eq:orga144570}
\text{s}\in\left\{ \text{bcc, fcc, hcp}\right\} \times\left\{ \gamma_{1},\gamma_{2},\gamma_{3}\right\}.
\end{equation}
The sum in \ref{eq:org13c10f9} is carried over all elements of the product from equation \ref{eq:orga144570}. The dominant interactions will be between first neighbours. With a single parameter decay, a set of three strains is the second smallest that resolves ambiguity arising from the dependence on the bond length. As for other parameters, \(\sigma\) indicates spin while \(A\) is a normalisation function that assigns higher weight to the majority spin. Additionally, the latter compensates for differences in number of eigenvalues in hcp bands. The sum is carried over all eigenvalues (index \(n\)) and samples of the Brillouin zone (index \(k\)). These details are relevant as they define the shape of the fitness function and an optimal solution as a result. This is due to fact that it is unlikely we will be able to achive a "perfect" fit. We distinguish between exponents \(p\) and \(p'\) so we can control the shape of a sphere with equally scored solutions and shape of the objective function separately. Other measures, such as the Jaccard similarity coefficient, Canberra distance \cmt{plot it and discuss symmetry??} or total variational distance, were also tested. However, the measure (\ref{eq:org13c10f9}) proved to be the most successful. It is flexible in terms of shaping of the fitness function landscape and can handle missing data or NaN values well. This is a critical property as some model parameters lead to an ill-posed eigenvalue problem. The summation always yields finite numbers, unless band calculations fail. Figure \ref{fig:org1d4418e} illustrates how changes in the value of \(p\) result in different "preferences" of the fitness function. Here, in agreement with Eq. \ref{eq:org13c10f9}, we assume that the fitness function is evaluated on two fixed points and we compare values of the candidate and reference functions.
The first row of plots in Figure \ref{fig:org1d4418e} shows top-right quarter of a unit sphere (for a given value of \(p\)) in two dimensions. It transpires that \(p\) norm \(1\) does not make a distinction between particular contributions to the distance. Fitness function will change by the same value, regardless how they are distributed. Higher values modify the sphere in a way that favours parallel deviations from a perfect fit. In other words, parallel deviations from the reference are more likely to be accepted if \(p\) is larger. Another way to observe this property is to investigate differences between vectors that are evenly distributed on the sphere (second row of Figure \ref{fig:org1d4418e}). These considerations are important in the context of fitting ill-posed problems as the best achievable solution will depend on the accepted fitness function (solutions based on different measures will approach the optimum differently).

\section{Defining a fitness measure for pair potentials}
\label{sec:org9b022ff}

To obtain a good estimate of an optimal pair-potential it is necessary to evaluate changes in total energy under different lattice deformations. Usually, this is done by evaluation of elastic constants. however, calculation of derivatives at a specific point introduces a certain level of ambiguity. Additionally, the values of elastic constants are extremely sensitive to even the smallest changes. For this reason evaluation of mechanical properties will use a similar measure to \ref{eq:org13c10f9}
\begin{equation}
\label{eq:orgb9ba81e}
f_{\text{pp}}\left(\vec{\alpha}_{\text{pp}}\right)=\left(\sum_{\boldsymbol{\eta}}\left|\frac{f_{\text{tb}}\left(\vec{\alpha}_{\text{pp}},\boldsymbol{\eta}\right)-f_{\text{ref.}}\left(\boldsymbol{\eta}\right)}{f_{\text{ref.}}\left(\boldsymbol{\eta}\right)}\right|^{p}\right)^{1/p'},
\end{equation}
where \(\vec{\alpha}_{\text{pp}}\) is a tuple of pair-potential parameters, \(\boldsymbol{\eta}\) is the strain tensor and \(f\) represents the elastic enthalpy for the tight-binding model (tb) and the reference (ref.). The measure (fitness function) for the pair-potentials has the distinct feature of being normalised to emphasise the importance of results near equilibrium crystal structure. At this point, we would like to distinguish an objective function from the fitness function. The objective function will be the target and a measure of "success", while the fitness function will be one that is used to drive the optimisation. The proposed approach relies on ab-initio (non-parametric) calculations, instead of on experimental values, although the latter are values we are aiming to recreate. However, in the case of ill-posed and non-linear problems sometimes by aiming towards a less optimal solution we can get "closer" to the objective. We note that in principle, it would be possible to recreate energy-strain curves using higher-order elastic constants. However, available measurements for temperatures close to absolute zero have an unspecified uncertainty and repeatability seems to be quite limited. This argument will be clarified in section \ref{sec:org7dd1231}. \cmt{Yes, it's about measurements. How to clarify this?}

\section{A summary of the selected fitness measures}
\label{sec:orgab9c557}

Deformations of a perfect crystal or one that contains defects should be sufficient to find all parameters of the model. However, these calculations are expensive because of the high dimensionality of the problem. On the other hand, we can evaluate the majority of parameters that are related to the electronic structure using fairly fast band-structure calculations. Furthermore, for evaluation of pair-potentials, we introduced a stable measure (less ambiguous and not overly sensitive) that includes information about higher-order elastic constants. Higher-order contributions will mainly depend on strains included in the fitness function. Finally, definitions \ref{eq:org13c10f9} and \ref{eq:orgb9ba81e} allow control over the family of acceptable solutions as well as the shape of the objective function (by controlling \(p\) and \(p'\) parameters). Furthermore, a combination of measures should provide an optimal amount of information with respect to computational time. Another way to look at these parameters is to consider \(p\) as controlling the penalty for a maximum difference between eigenvalues across the Brillouin zone, while \(p'\) controls the penalty for poorly chosen model parameters.

\section{Selecting objectives}
\label{sec:org27221d6}

\begin{figure}[htbp!]

\includegraphics[height=0.5\textwidth]{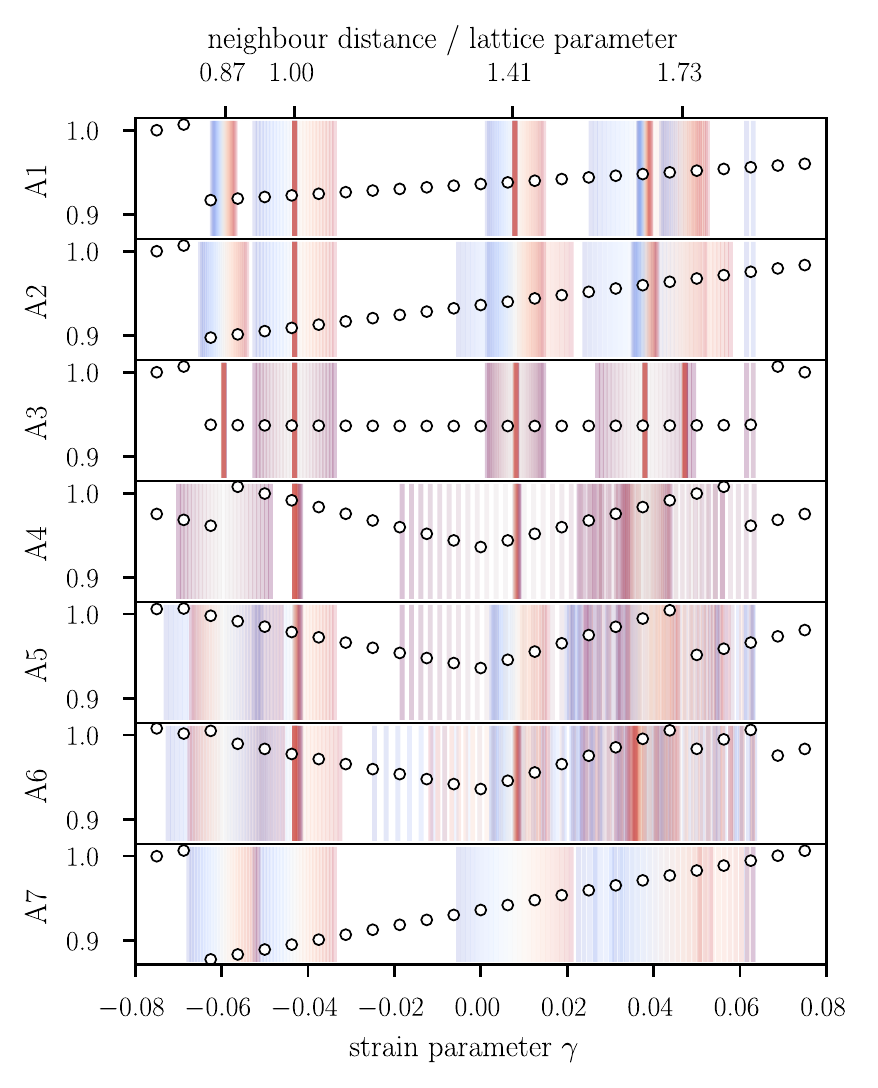}
\caption{\label{fig:org69a45cf}Evolution of distances, measured on the upper horizontal axis, between pairs of atoms (vertical lines) under deformations of a bcc lattice presented in Table \ref{tab:orgf2d28c0} (excluding deformation A8). Blue colour indicates negative values of \(\gamma\) while red - positive. Scatter plot represents normalised maximum distance included within cut-off radius \(2 a_{\text{bcc}}\), where \(a_{\text{bcc}}\) is the equilibrium bcc lattice parameter (\(2.866\, (\text{\AA})\)). \cmt{not true anymore-- Deformation patterns A1-A6 are necessary to calculate second and third order elastic constants. Pattern A7 (tension-compression) is used in fitting of bands}.}
\end{figure}

In this section, the objective is to define a set of model predictions that will be combined using the above measures to give a numerical representation of the fitness. To provide some overall context, consider a set of deformations necessary to represent higher-order elastic constants of cubic materials. Considered deformation patterns are given in Table \ref{tab:orgf2d28c0}. 
\begin{table}[htbp!]
\caption{\label{tab:orgf2d28c0}Components of the Voigt strain used to construct deformation tensors that defines lattice deformations.}
\centering
\small
\begin{tabular}{llrrrrr}
\hline
pattern / component & \(\eta_{1}\) & \(\eta_{2}\) & \(\eta_{3}\) & \(\eta_{4}\) & \(\eta_{5}\) & \(\eta_{6}\)\\
\hline
A1 & \(\gamma\) & 0 & 0 & 0 & 0 & 0\\
A2 & \(\gamma\) & \(\gamma\) & 0 & 0 & 0 & 0\\
A3 & \(\gamma\) & \(-\gamma\) & 0 & 0 & 0 & 0\\
A4 & 0 & 0 & 0 & 0 & 0 & \(2 \gamma\)\\
A5 & \(\gamma\) & 0 & 0 & \(2 \gamma\) & 0 & 0\\
A6 & \(\gamma\) & 0 & 0 & 0 & \(2\gamma\) & 0\\
A7 & \(\gamma\) & \(\gamma\) & \(\gamma\) & 0 & 0 & 0\\
A8 & 0 & 0 & 0 & \(2\gamma\) & \(2\gamma\) & \(2\gamma\)\\
\hline
\end{tabular}
\end{table}
Patterns A1-A6 are based on ones given in \cite{KhVekilov2016}. Additional patterns A7 and A8 are necessary to resolve \(c_{123}\) and \(c_{456}\) elastic constants. Explanation for our selection of objectives we begin by arguing that, deformations necessary to calculate higher order elastic constants, suffice to sample the space of local atomic environments appropriately. That is, by aiming to recreate elastic constants (up an arbitrary order and by including various crystallographic structures) it is possible to resolve all model parameters. Here it is just an assumption. However, we can gain some confidance in this statement by looking at the influence of various deformations on the sampling of the pair-wise interactions (Figure \ref{fig:org69a45cf}).

However, while deformations in table \ref{tab:orgf2d28c0} are necessary to estimate elastic constants, some of them are not necessary the best choice for evaluation of how well a model can represent bands structures. This assertion is based on our experience and we argue that this is a result of emphasis on further neighbours. For example, taking deformations A1 and A3 will provide us with test cases that are different only in minor features of a band structure when model parameters are optimal. Furthermore, it is likely that we wont be able to recreate them due to model limitations. These considerations are important as evaluation of bands is an inexpensive source of high quality information about the goodness of fit. Hence, we would like to put more emphasis on these calculations in the fitting process. Therefore we shift our focus to the pattern A7. Deformations A1-A6 will again become relevant in the context of elastic properties. We would like to implicitly sample the relation between distance and magnitude, for hoppings and overlaps alike, across all relevant distances between pairs of orbitals. As we can see from Figure \ref{fig:org69a45cf}, deformations vary significantly in the way that they affect the distribution of distances between pairs. On the other hand, the starting point for the model is non-self-consistent TB, a method that can be at best a variational approximation to the ground state, where effective potential/density is implicitly defined by hopping parameters. Therefore, the primary objective should be a recreation of band structures at equilibrium lattice parameters for realistic structures. Hence, we select test cases that maximise sensitivity to changes in parameters while not deviating too much from the ground state. By emphasising the variety of structures over a variety of deformations we are also minimising the issue of not being able to recreate changes in the reference as higher strains may result in changing the number of neighbours within the cut-off radius. We found that even in case of a simple compression or tension it is impossible to force TB to recreate a certain class of changes in band structures that DFT or QSGW predicts. Selection of the crystallographic structures is based on the information provided in the Table \ref{tab:orgcaf7081}.
\begin{table}[htbp]
\caption{\label{tab:orgcaf7081}Population size of neighbour-shells in different structures.}
\small
\begin{tabular}{lrrr}
\hline
neighbour/structure & 1st & 2nd & 3rd\\
\hline
diamond & 4 & 12 & 12\\
simple cubic & 6 & 12 & 8\\
bcc & 8 & 6 & 12\\
hcp & 12 & 6 & 2\\
fcc & 12 & 6 & 24\\
\hline
\end{tabular}
\end{table}
The most natural candidates are close-packed structures: hcp and fcc. These are known to work well with the atomic-sphere approximation (ASA) making them an "easy target" for TB as TB and ASA are very closely related \cite{QeASA,Andersen1984} \cmt{Still not sure about this arg.}. Needless to say the bcc structure needs also to be included in the reference data as we are optimising the model of ferromagnetic Fe. Due to the similarity between fcc and hcp, namely the same number and distances of 1st and 2nd neighbour-shells, there is a danger of bias in the reference data-set (over-fitting). Other structures included in Table \ref{tab:orgcaf7081} show potential improvements in the reference. Regardless of the selected structures we also need to include strained structures as three structures will be insufficient to estimate optimal dependence of pair-wise interactions on distance. We decided to focus on the easiest to recreate dependence on isotropic compression and tension. As illustrated in Figure \ref{fig:org69a45cf} (A7), this should provide a sufficiently rich sampling. However, we accepted this simple reference set only because decay is defined by a single parameter. A more flexible model might require a more complex objective/fitness function. The final consideration in this scope is the magnitude of strain. The plot of the maximum pair-wise distance within cut-off radius under volumetric strain (A7) demonstrates that under high strains we change the number of orbitals included in the construction of the Hamiltonian. This is indicated by the sudden increase of the maximum distance under compression. Hence, effectively we change the nature of the model leaving us with no other choice other than limit the strain magnitude. Otherwise, we push the optimal solution further away from what can be achieved within the TB approximation. This is one of the reasons we limit ourselves to the following volumetric strains: -0.06, 0.00, 0.06. This gives us \(3\times3\) structures included in the fitness function \ref{eq:org13c10f9}. There is also another reason for this approach that will be addressed in section \ref{sec:org20e90c2}.

The next piece of the puzzle is the recreation of the mechanical properties. Here we address why we decided to aim for deformations that are representative for higher elastic constants. In our experiments, we noticed that any subset of 3 patterns from Table \ref{tab:orgf2d28c0} can be used to calculate second-order elastic constants (SOEC). However, that does not guarantee that the bcc structure will be stabilised. From Figure \ref{fig:org69a45cf} we can see that mode A3, that explicitly gives us the \(c'\) shearing constant, emphasises other than nearest-neighbour (NN) distances. It is possible to obtain a very good estimate of elastic constants, as a result of cancelling errors, by overestimation of interactions at NN. In many cases, this could mean a negative curvature of the energy-strain relation when using the A3 deformation pattern.

\section{Optimisation strategy}
\label{sec:org20e90c2}
\cmt{Normally we would constrain as much as possible but it's ill-posed so we prefer flexibility.} While the objective is to predict total energies associated with defects, we argue that these calculations are too expensive to be included in the reference data-set. Therefore, we focus on the measures presented above, associated with two features of the model predictions: electronic structure and elastic properties. It is likely that given a sufficiently large reference set all ambiguities can be resolved. However, calculations of elastic properties can also be very expensive given the nature of the problem (non-linearity and with high-dimensionality). Furthermore, it might be impossible to assemble all results within a single objective without arbitrarily selected weighting. For example, different elastic constants can vary by orders of magnitude and cannot collectively be compared with formation energies. Here, multi-objective optimisation is out of the question because it is simply too expensive. Therefore, we make a case for the classical approach where optimisation is based on separate, also inexpensive, band calculations and estimation of elastic constants (energy-strain curves in our case). Equation \ref{eq:orgdb9beb0} shows that to a degree the optimisation problem is separable. The drawback of this approach is that a good fit to the band structures does not necessarily indicate that it will be possible to find a reasonable approximation to the optimal pair-potential (leading to poorly recreated elastic constants). On the other hand, we would like to avoid situations in which good candidates for hopping integrals are rejected in an optimisation step because the algorithm is exploring the wrong subspace of pair-potential parameters at the time. Nonetheless, it is necessary to combine knowledge about both sub-sets of parameters at some point of the optimisation. We have solved this problem by using a brute force approach. The idea is to generate a large sample of candidate band-structure parameters. For that reason, we optimised parameters in several stages using different p-norms and references. Parameters were fitted against DFT band-structures with LDA and PBE functionals as well as the QSGW method \cmt{plot?}. Each reference was evaluated using \(p \in \{1,2,3\}\). Subsequently, from every population (9 in total), three samples were selected based on performance with respect to each measure. This gave us 27 candidates from a single run. Starting from the previous optimum this procedure was repeated several times, giving more than 100 candidate parameterisations. \cmt{Make a flow-chart? Limited added value though.}
%% \begin{tikzpicture}[node distance=2cm]
%% \tikzstyle{startstop} = [rectangle, rounded corners, minimum width=.5cm, minimum height=.5cm,text centered, draw=black, fill=red!30]
%% \tikzstyle{process} = [rectangle, minimum width=3cm, minimum height=1cm, text centered, draw=black, fill=orange!30]
%% \node (start) [startstop] {Start};
%% \end{tikzpicture}

\begin{figure}[htbp!]
\centering
\includegraphics[width=0.4\textwidth]{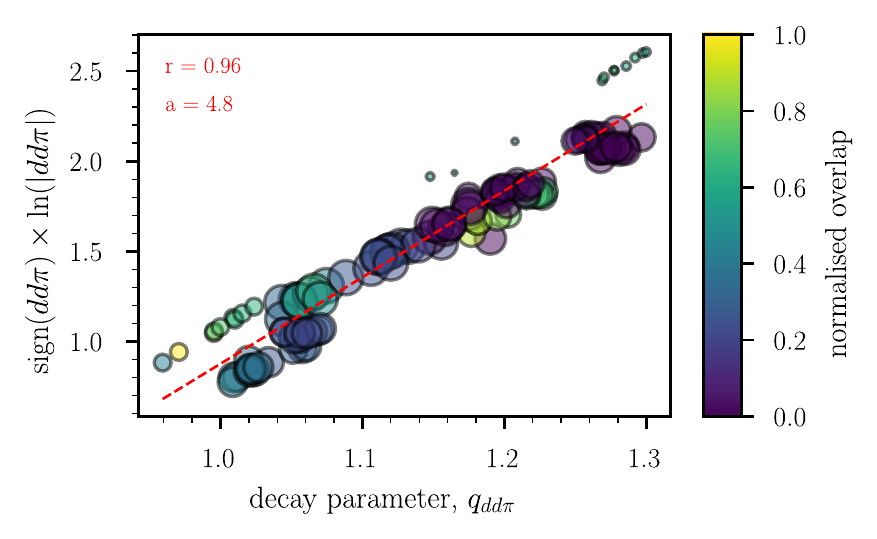}
\caption{\label{fig:org2b86ef6}Example of a relationship between the fitness function value, magnitude, decay and overlap. Given the exponential decay, the majority of other hopping integrals revealed a similar relationship. Size of the data-points is proportional to the value of the fitness function (bigger marker indicates a better performance). The red line represents a linear fit with a value of the slope given as the variable \(a\) and correlation coefficient as \(r\). Colours represents a scaled overlap that in case of the \(d\text{-block}\) could be considered as relatively small.}
\end{figure}

\begin{figure*}[htb!]
\centering
\includegraphics[width=.9\linewidth]{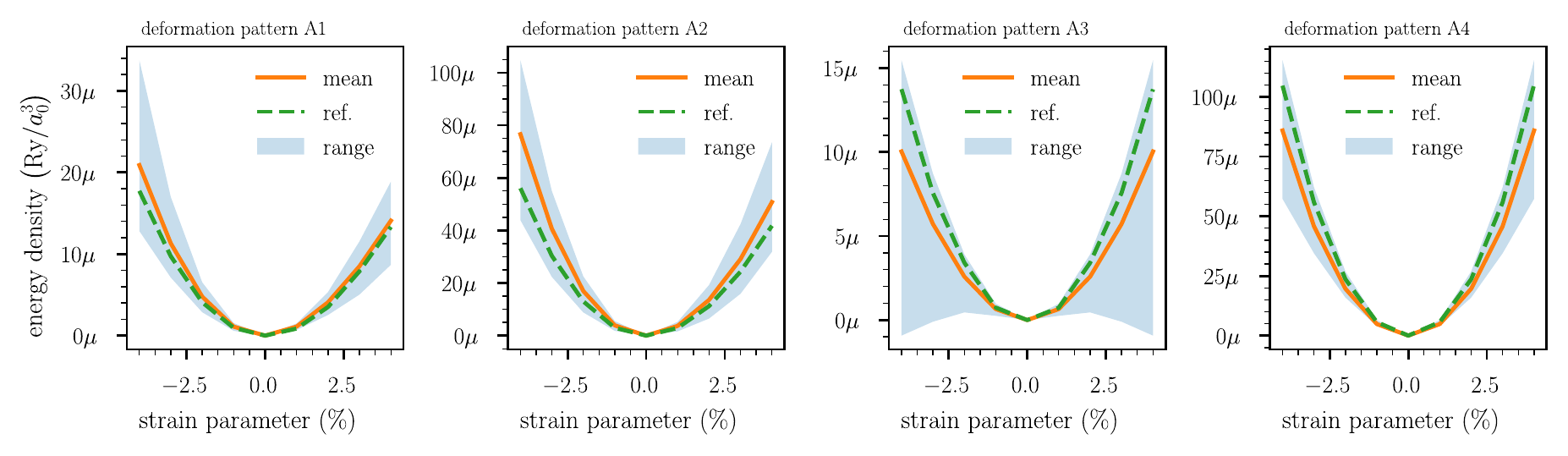}
\caption{\label{fig:org3866ec6}The mean and range of energy-strain relationships for selected deformation patterns from the reference-data-set as estimated by candidate tight-binding models. Note the negative values for the deformation pattern A3.}
\end{figure*}

In our experiments, we tested several optimisation algorithms focusing on derivative-free methods. We found that the covariance matrix adaptation evolution strategy (CMAES, \cite{Hansen1996}) was the most efficient and robust. Particle swarm optimisation (PSO) also performed well. Surprisingly, the Gaussian process optimisation with radial kernels failed to find any reasonable solution. We decided, to focus on the CMAES using an implementation by Hansen et al. \cite{hansen2019pycma}. Broadly speaking, in the CMAES a population of random vectors (smaples) in the space of model parameters is generated from a multivariate Gaussian distribution defined by the initial mean and covariance. Subsequently, each sample is evaluated and ranked according to the fitness function. On this basis, the mean and covariance matrix of the distribution is updated and the cycle is repeated. The rule-of-thumb was to set the size of the population to 10-15 times the number of degrees of freedom. The optimisation would not be successful without setting some initial constraints on the parameters. This was implemented by forming 3 groups of parameters: 1: \(\{dd\sigma, dd\pi, dd\delta\}\) (d-block), 2: \(\{sd\sigma, pd\sigma, pd\pi\}\) (sp-d block), 3: \(\{ss\sigma, sp\sigma, pp\sigma, pp\pi\}\) (sp block). The partition was inspired by the approach by Dufrense et al. presented in \cite{Dufresne2015}. Each group initially shares the decay parameters and overlaps. We placed no constraints on ratios between hopping integrals. Following convergence of the fitness function, we recorded the results as described earlier. \cmt{Plot of convergence?} In the next optimisation step, we modified the formulation of the problem, allowing, for example, each decay parameter to be optimised separately. The constraints were imposed and removed in each cycle after a careful examination of the features of the band structures. The idea was to build a sample of candidates that successfully recreated band structures. \cmt{Show population?} At this stage, we found only basic relationships between parameters. An example can be found in Figure \ref{fig:org2b86ef6}.
From this figure, it is clear that there is a strong correlation between the magnitude and decay amongst optimised parameters. Furthermore, there is a significant level of ambiguity between these quantities. This indicates that the value of the hopping integral at nearest-neighbours is a dominant factor here. Likewise, overlap parameters are ambiguous. Our approach allows us to explore a wide variety of potential candidates in subsequent tests of other properties (e.g. elastic const.). However, there is no guarantee that it will be possible to find a unique solution using computationally inexpensive tests. After several cycles of optimisations with different measures and reference band structures, we attempted to fit energy-strain curves and evaluate elastic constants. To speed up the process, we precalculated energies without pair potentials for each deformation pattern. In the optimisation of pair-potentials, we used the \(p=3\) norm and recently developed SHGO algorithm, developed by Endres et al. \cite{Endres2018}, with Sobol sampling, implemented in the SciPy stack \cite{2019arXiv190710121V}. We emphasise that we tested a variety of other algorithms and all failed to provide any acceptable results. The main issue was negative curvature in deformation pattern A3. \cmt{Plot?} A very similar problem was encountered by Mehl et al. \cite{Mehl1996}. Though this also occurred with SHGO, nonetheless, this algorithm allowed us to find a satisfactory solution with average relative differences in total energy below 20\% for all tested deformation patterns. Figure \ref{fig:org3866ec6} illustrates the problem.
In principle, it would be possible to estimate elastic constants ignoring the pattern A3 and obtain satisfactory predictions. The distinct feature of this deformation pattern is weak dependence on the 1-st neighbour shell (Figure \ref{fig:org69a45cf}). We argue that by including deformations that allow calculation of higher-order elastic constants we introduced a more robust framework. The drawback is that we need to fit energy-strain curves. Hence, we cannot rely on experimental data. At this point, we arrived at what seemed to be a Pareto front between the good representation of band structures and of elastic constants. This means further improvements in band structure resulted in worsening of the mechanical properties and vice versa. To verify that is the case, we compared hopping and overlap parameters that gave best band structures with a subset that resulted in the best mechanical properties. We used upper quantile planes from both sets and by taking the intersection approximated a region of parameter space that most likely will satisfy both requirements. Details of this approach will be published separately. This was an essential step to obtain the parameters presented in Section \ref{sec:org201a625}. Having placed a new set of bounds on our model parameters we repeated the previous procedure with several further cycles of optimisation, this resulted in noticeable improvements. From the new set of results, we selected the most promising as a new starting point and changed the optimisation method. This time multi-stage local optimisation was applied. At each stage, we targeted only the bcc band structures for the majority spin with the Nelder-Mead method \cite{2019arXiv190710121V,Gao2012,Nelder1965}. This way we had a much bigger chance of achieving a near-perfect fit and optimised the very subtle features of the band structure that are shaped by hopping integrals associated with p-orbitals. In each stage of the optimisation, we selected a different search direction in the parameter space. For example, first, we optimised ratios for d-band parameters, then scales for each group of hopping integrals, together with the Stoner parameter and so on. Globally, we employed a semi-automatic framework where most of the steps were automated. However, the order and form of constraints would depend on the success of the optimisation. This way a sample of approximately 200 models was created. The three most promising candidates will be presented and evaluated in the following sections.

\section{Parameters of the electronic structure}
\label{sec:org201a625}
\begin{figure*}[htbp!]
\begin{center}
\includegraphics[width=.9\linewidth]{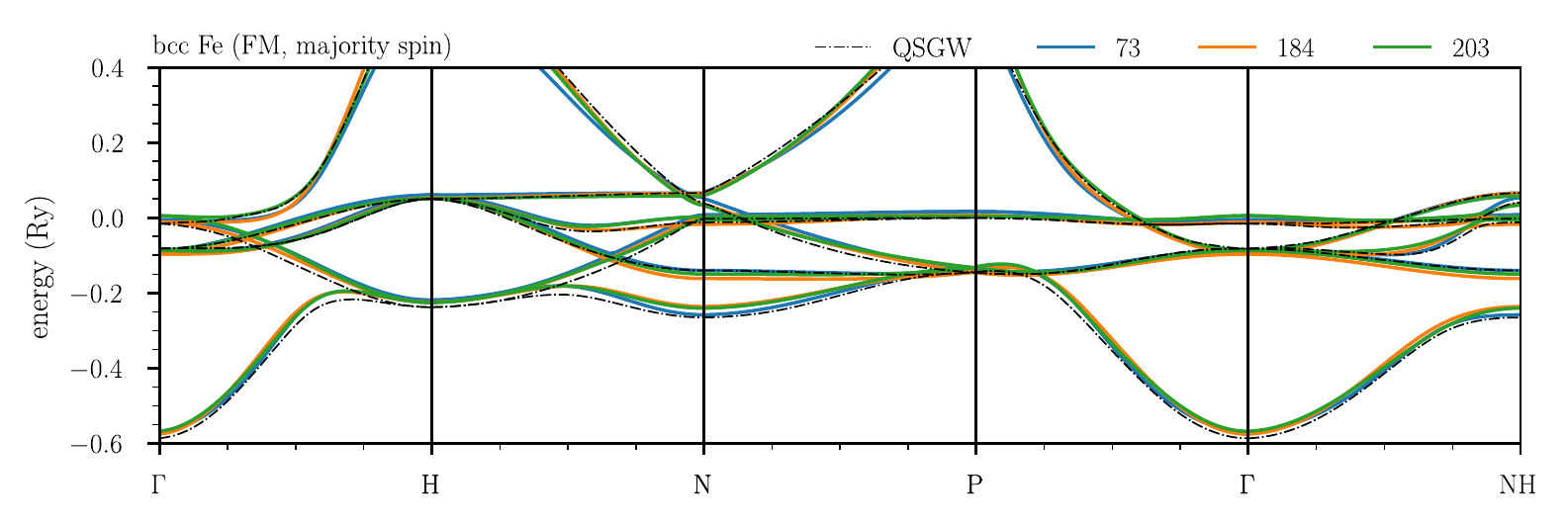}
\end{center}
\begin{center}
\includegraphics[width=.9\linewidth]{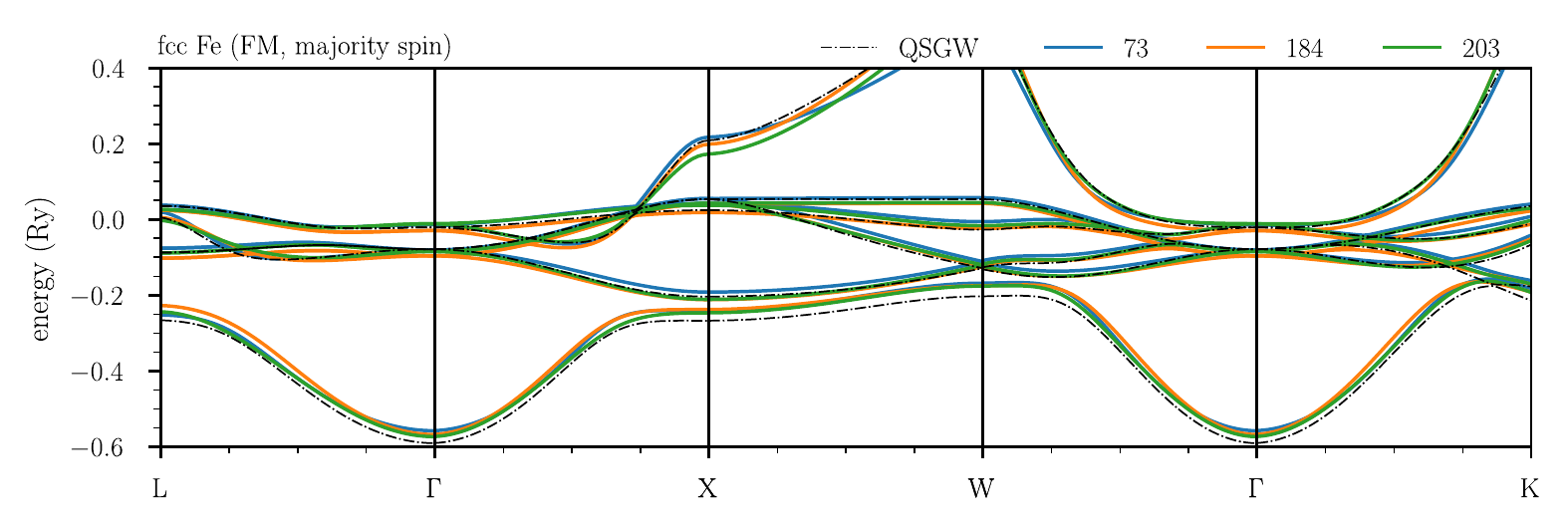}
\end{center}
\begin{center}
\includegraphics[width=.9\linewidth]{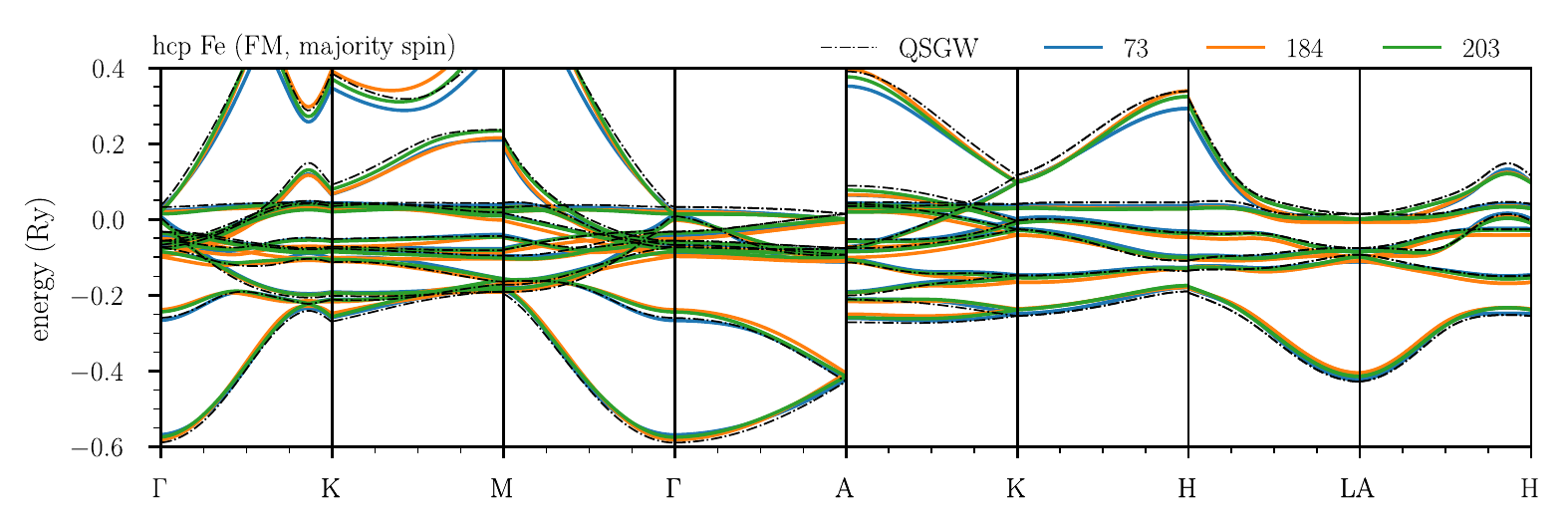}
\end{center}
\caption{Band structures for bcc, fcc and hcp structures of Fe in the ferromagnetic state. Here, the fcc lattice is under expansion 2\% and .35\% for models 73 and 184 respectively. The black dotted line represents band structures plotted from electron density updated in the QSGW self-consistency cycle. \cmt{Consider zooming in the gamma point?}}
\label{fig:bnds_final}
\end{figure*}

\begin{figure*}[htbp!]
\centering
\includegraphics[width=.9\linewidth]{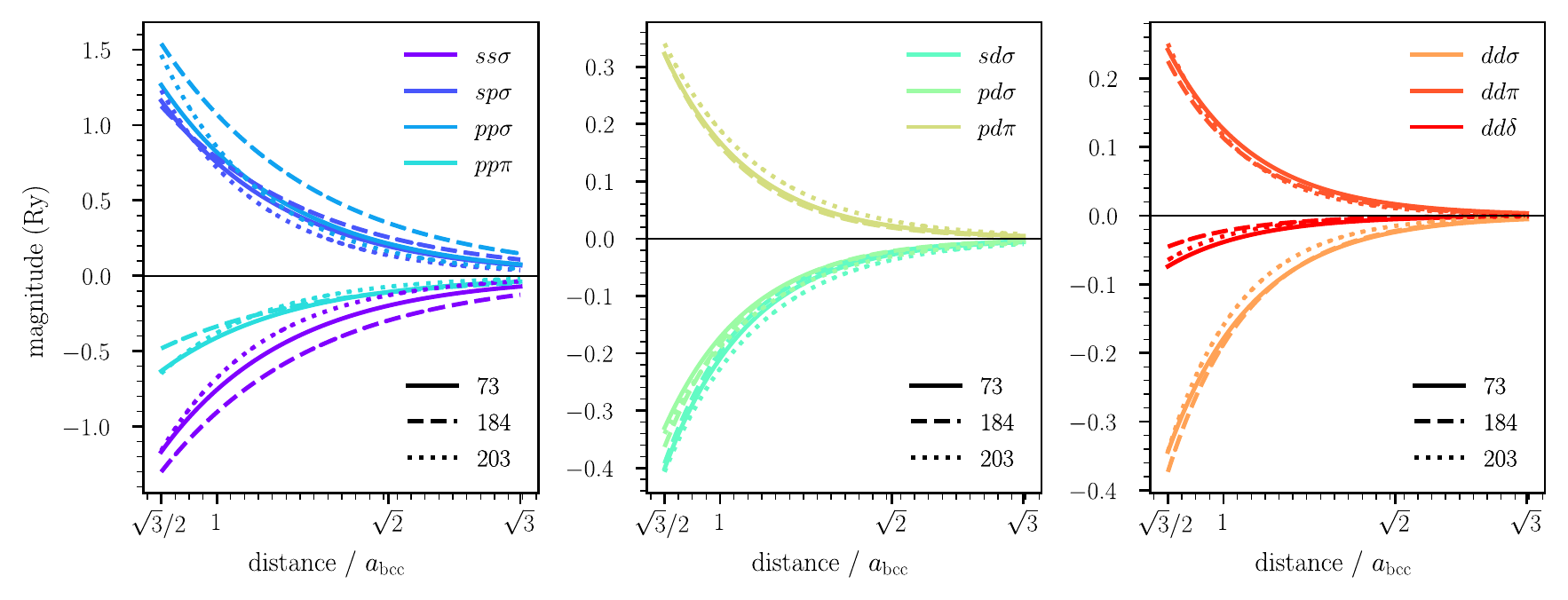}
\caption{\label{fig:orge1ba165}Nominal values (excluding augmentation) of hopping integrals as a function of distance between orbitals. Ticks indicate subsequent neighbour shells in the bcc structure.}
\end{figure*}

For the sake of reproducibility, we now present details of electronic structure parametrisation. The fundamental relation between hopping magnitude and distance \(|\vec{r}_{\alpha\beta}|\) between orbitals \(\alpha\) and \(\beta\), is given by an exponential decay i.e.
\begin{equation}
\tilde{h}_{\alpha\beta}(\vec{r}_{\alpha\beta}) = 
m_{\alpha\beta} \, e^{-q_{\alpha\beta}\, |\vec{r}_{\alpha\beta}|},
\end{equation}
where \(m\) controls the magnitude and \(q\) is the decay parameter. The hopping function \(h_{\alpha\beta}(\vec{r}_{\alpha\beta})\) used in the calculation of Bloch sums and construction of the Hamiltonian is simply
\begin{equation}
\label{eq:org9a6cd78}
h_{\alpha\beta}(\vec{r}_{\alpha\beta}) = 
\tilde{h}_{\alpha\beta}(\vec{r}_{\alpha\beta})
\end{equation}
for \(|\vec{r}_{\alpha\beta}|\) smaller than the radius \(r_\text{A}\). Otherwise,
\begin{equation}
\label{eq:org3de080d}
h_{\alpha\beta}(\vec{r}_{\alpha\beta}) = 
\begin{cases}
P_5\left(\tilde{h}_{\alpha\beta},r_{\mathrm{A}},r_{\mathrm{C}}\right) & \text{agm.} \\
\tilde{h}_{\alpha\beta}(\vec{r}_{\alpha\beta})\,\tilde{P}_5\left(\tilde{h}_{\alpha\beta},r_{\mathrm{A}},r_{\mathrm{C}}\right) & \text{multi.} \\
\end{cases},
\end{equation}
depending on the method of control of the tails of hopping integrals. The first case (agm) denotes augmentative cutoff where, \(\tilde{h}_{\alpha\beta}\)  is replaced by a fifth-order polynomial \(P_5\) that matches value, slope and curvature of \(\tilde{h}_{\alpha\beta}\) at \(r_\text{A}\), while setting all three to zero at the critical radius \(r_\text{C}\) and beyond.  In the second case of multiplicative cutoff (multi.) \(\tilde{P}_5\) is a very similar polynomial with the exception of being normalised so it evaluates to 1 at \(r_\text{A}\). In the basis set we include \(3d\), \(4s\) and \(4p\) orbitals. Furthermore, the Stoner parameter \(I\) was the same for all l-channels and the Hubbard-like \(U\) was set to \(1 \, \mathrm{Ry}\). In this regard, we follow the model of Paxton et al. \cite{Paxton2008}. The most promising models presented here are ones from the \([\text{GSGW}, p_1, p_1]\) group. This means that the references were bands from QSGW calculations, the fitness function was normalised by \(p=1\) and also the score used to select the best sample from the optimised population was calculated in the same way as the fitness. Each model is identified by a code from our data-set namely 73, 184 and 203. The resulting band structures are plotted against the reference and presented in Figure \ref{fig:bnds_final}.
We focus on the ferromagnetic phase for all structures as this one was the reference. Note that in initial optimisation of all three band structures the Stoner parameter was much higher. It was reduced later to obtain near-perfect fit. It can immediately be seen from Figure \ref{fig:bnds_final} that we obtained a very good agreement with the reference. Results for the fcc structure were calculated under strain as two TB models do not predict the meta-stable ferromagnetic state if volume per atom matches the stable bcc lattice. Thanks to the \(spd\) basis we are able to recreate all the features up to energy levels that do not involve \(4d\) orbitals. Although these levels are not shown here, in two cases we were able to maintain band topology in the case of the bcc structure (models 184 and 203). For this reason, these models have a shorter cut-off on the \(sp\) block of parameters. Needless to say, bands were tested under strain and they maintain most of the features, mainly for eigenvalues that do not involve p-orbitals. Values of all parameters associated with the electronic structure are presented in Table \ref{tab:hi}.
\begin{table*}[htb!]
\small
Off-diagonal hopping parameters\\
\begin{tabular}{lrrrrrrrrrrrr}
\toprule
{} & \multicolumn{6}{r}{hopping} & \multicolumn{6}{r}{overlap} \\
\cmidrule(lr){2-7}\cmidrule(lr){8-13}
{} & \multicolumn{3}{r}{magnitude (Ry)} & \multicolumn{3}{r}{decay ($\mathrm{a}^{-1}$)} & \multicolumn{3}{r}{magnitude (1)} & \multicolumn{3}{r}{decay ($\mathrm{a}_0^{-1}$)} \\
\cmidrule(lr){2-4}\cmidrule(lr){5-7}\cmidrule(lr){8-10}\cmidrule(lr){11-13}
{} &            73  &    184 &    203 &                       73  &   184 &   203 &           73  &    184 &    203 &                         73  &   184 &   203 \\
\midrule
$ss\sigma$ &         -1.935 & -1.350 & -3.740 &                     0.600 & 0.500 & 0.742 &         1.984 &  1.160 &  2.500 &                       0.600 & 0.500 & 0.742 \\
$sp\sigma$ &          1.927 &  1.169 &  3.974 &                     0.600 & 0.500 & 0.742 &        -2.200 & -1.142 & -3.500 &                       0.600 & 0.500 & 0.742 \\
$pp\sigma$ &          2.100 &  1.600 &  4.730 &                     0.600 & 0.500 & 0.742 &        -2.809 & -0.950 & -4.200 &                       0.600 & 0.500 & 0.742 \\
$pp\pi$    &         -1.048 & -0.500 & -2.090 &                     0.600 & 0.500 & 0.742 &         0.350 &  0.143 &  0.150 &                       0.600 & 0.500 & 0.742 \\
$sd\sigma$ &         -2.711 & -3.349 & -1.783 &                     0.900 & 0.950 & 0.808 &         0.000 &  0.000 &  0.636 &                       0.000 & 0.000 & 0.808 \\
$pd\sigma$ &         -2.232 & -3.104 & -1.492 &                     0.900 & 0.950 & 0.808 &         0.000 &  0.000 &  0.710 &                       0.000 & 0.000 & 0.808 \\
$pd\pi$    &          2.179 &  2.754 &  1.492 &                     0.900 & 0.950 & 0.808 &         0.000 &  0.000 & -0.658 &                       0.000 & 0.000 & 0.808 \\
$dd\sigma$ &         -2.322 & -3.186 & -4.874 &                     0.900 & 0.950 & 1.058 &         0.000 &  0.000 &  0.000 &                       0.000 & 0.000 & 0.000 \\
$dd\pi$    &          1.633 &  1.922 &  3.555 &                     0.900 & 0.950 & 1.058 &         0.000 &  0.000 &  0.000 &                       0.000 & 0.000 & 0.000 \\
$dd\delta$ &         -0.499 & -0.386 & -0.917 &                     0.900 & 0.950 & 1.058 &         0.000 &  0.000 &  0.000 &                       0.000 & 0.000 & 0.000 \\
\bottomrule
\end{tabular}

\phantom{} \medskip  On-site, cut-offs and Stoner parameter \\
\begin{tabular}{lrrrrrrrrrr}
\toprule
{} &  $\epsilon_{3d} (\mathrm{Ry})$ &  $\epsilon_{4s} (\mathrm{Ry})$ &  $\epsilon_{4p} (\mathrm{Ry})$ &  $r_{\mathrm{A1}} (\mathrm{a}_{0})$ &  $r_{\mathrm{A2}} (\mathrm{a}_{0})$ &  $r_{\mathrm{A3}} (\mathrm{a}_{0})$ &  $r_{\mathrm{C1}} (\mathrm{a}_{0})$ &  $r_{\mathrm{C2}} (\mathrm{a}_{0})$ &  $r_{\mathrm{C3}} (\mathrm{a}_{0})$ &   $I$ \\
\midrule
73  &                         -0.002 &                          0.248 &                          0.718 &                               8.504 &                               8.504 &                               8.504 &                              10.087 &                              10.087 &                              10.087 & 0.046 \\
184 &                         -0.004 &                          0.346 &                          0.843 &                               8.504 &                               8.504 &                               6.525 &                              10.087 &                              10.087 &                               8.504 & 0.049 \\
203 &                          0.000 &                          0.350 &                          0.830 &                               8.504 &                               8.504 &                               6.525 &                              10.087 &                              10.087 &                               8.504 & 0.049 \\
\bottomrule
\end{tabular}

\caption{Table of tight-binding parameters of Fe associated with the electronic structure. The Hubbard-like U is $1 \, \mathrm{Ry}$ in all cases. Orbtails included in the model are naturally $3d$, $4s$ and $4p$}.
\label{tab:hi}
\end{table*}
The biggest difference between models is the decay parameters in the \(sp\) block as these were the parameters that were modified to obtain better mechanical properties in the last (semi-automatic) optimisation stage. However, in a set of over 200 models, no significant correlations (usually below 30\%) between these quantities were found. Hence, at best, this step should be considered as a method to explore the subspace of parameters that provide a very good agreement in terms of band structures. From this subset, we can select ones that also allow us to find good elastic constants. In the future, we plan to find a way to find a relation between parameters within this region and optimal mechanical properties. As mentioned before, the cut-off radius was modified in the case of models 184 and 203 to obtain a better representation of high-energy molecular orbitals in the bcc structure. Furthermore, there was a strong preference towards an orthogonal sp-d block. Therefore, in the case of model 184, this group of orbitals was explicitly set to be orthogonal. Apparent variations in other parameters might be misleading due to ambiguity arising from accepting hopping magnitude as a function of bond length. This aspect of the models is illustrated by a logarithmic relationship between magnitude and decay (see e.g. Figure \ref{fig:orge1ba165}).
It is clear that all estimates of optimal hopping integrals that involve \(d\) orbitals are very similar. Although, these differences are sufficient to influence the mechanical properties. Not surprisingly, the biggest variations can be observed in the case of \(sp\) orbitals.
\section{Elastic properties}
\label{sec:org7dd1231}
The models presented in this paper are based on two types of repulsive pair-potentials -- \(V\). The first is referred to as Goodwin-Skinner-Pettifor (GSP, \cite{Goodwin1989}) and has the form
\begin{equation}
\label{eq:org00f58d7}
V_{\text{GSP}}(\vec{r}) = 
A \left(\frac{r_0}{r} \right)^{n} 
\exp \left( \left( \frac{r}{r_\text{c}} \right)^{n_\text{c}}  - \left( \frac{r_0}{r_\text{c}} \right)^{n_c} \right)^{n},
\end{equation}
where \(r=|\vec{r}|\) and \(n\), \(n_c\), \(V\), \(r_0\) and \(r_c\) are adjustable parameters, each controlling different features of the potential. Here, we apply a shortened notation where dependence on the parameters is implicit. The second type of potential has a similar form to that for the  hopping integrals i.e.
\begin{equation}
\label{eq:org2bbf434}
V_{\text{exp}} (\vec{r}) = \left(\sum_{i=1}^{3} a_i r^{b_i} \exp \left( -c_i r \right) \right),
\end{equation}
where for evaluation of values of the potential we use the same augmentation procedure as in the case of hooping integrals and overlap functions (Equation \ref{eq:org3de080d}). Hence,
\begin{equation}
V(\vec{r}) = V_\text{exp/GSP}(\vec{r}) P_5 \left(V_\text{exp/GSP}, r_\text{A},r_\text{C} \right).
\end{equation}
Here we distinguish between \(r_\text{c}\) and \(r_\text{C}\). The former is a internal parameter of \(V_\text{GSP}\) function while the latter is a global cut-off parameter that defines which interactions will be set explicitly to zero. Potential \(V_\text{GSP}\) does not need to be augmented. Nonetheless, the code was set to augment this potential right before \(r_\text{C}\). Furthermore, the value of \(r_\text{C}\) was set to \(2a_\text{bcc}\), where \(a_\text{bcc}\) is the equilibrium-volume lattice parameter. When using \(V_\text{exp}\) the augmentation becomes a critical factor and both associated parameters, \(r_\mathrm{A}\) and \(r_\mathrm{C}\), were also subject to optimisation.

\begin{figure*}[htb!]
\centering
\includegraphics[width=.9\linewidth]{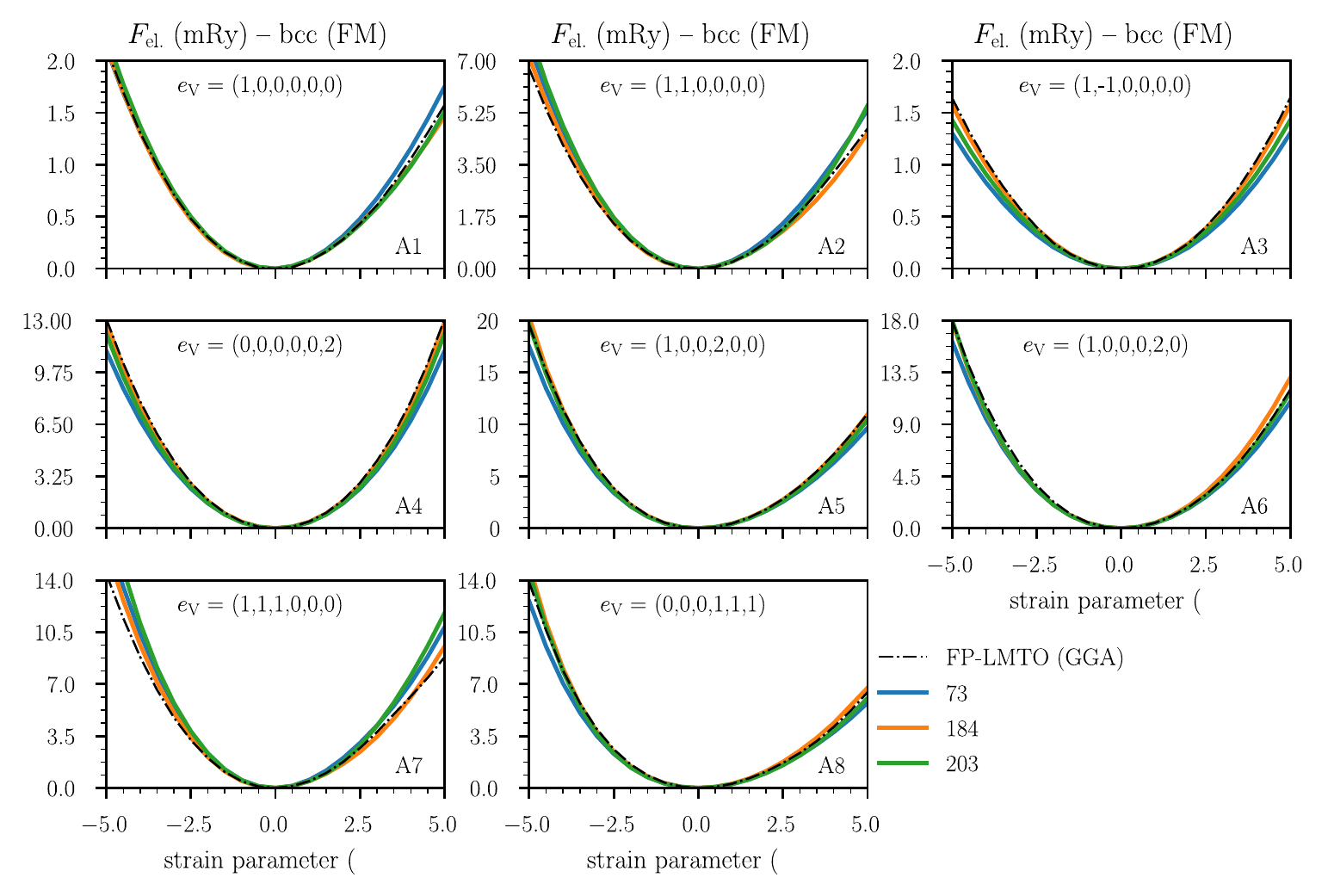}
\caption{\label{fig:org3b00bf4}Energy-strain curves for 8 different strain patterns. Energy is normalised by the number of atoms in a unit cell. Strain patterns are given in Voigt notation. Piece-wise linear interpolation was used and the distance between data-points is 0.5\%.}
\end{figure*}

In optimisation of the pair-potentials, we could not rely on QSGW calculations as a reference. As mentioned above, we instead used FP-LMTO with the PBE91 exchange-correlation functional \cite{Pashov2019,pbe1996}. The objective was to recreate energy-strain curves in deformations defined in Table \ref{tab:orgf2d28c0}. The results are presented in Figure \ref{fig:org3b00bf4}.
Estimates of optimal parameters are given in Table \ref{tab:pp}.
\begin{table}[htb!]
\small
Model with $V_{\text{GSP}}$ pair-potiential\\
\begin{tabular}{lrrrrrrr}
\toprule
{} &  $r_{\text{A}}$ &  $r_{\text{C}}$ &      $A$ &    $m$ &  $m_{\text{c}}$ &  $r_{0}$ &  $r_{\text{c}}$ \\
\midrule
\textbf{73} &          10.757 &          10.811 &  931.102 &  7.481 &          10.567 &    1.312 &           6.337 \\
\bottomrule
\end{tabular}

\phantom{} \medskip  Models with $V_{\text{exp.}}$ pair-potential \\
\begin{tabular}{lrrrrr}
\toprule
{} &  $r_{\text{A}}$ &  $r_{\text{C}}$ &     $a_1$ &  $b_1$ &  $c_1$ \\
\midrule
\textbf{184} &           4.353 &           6.493 &  1399.534 & -1.037 &  1.817 \\
\textbf{203} &           4.379 &           6.832 &   768.104 & -0.000 &  2.004 \\
\bottomrule
\end{tabular}

\caption{Parameters of pair-potentials. Parameters not specified in lower table are set to 0. \cmt{0.5 day of work to plot them, with augmentation}}
\label{tab:pp}
\end{table}
It immediately transpires that we were able to obtain a good agreement with the reference. The tendency to underestimate total energy in case of A3 is preferable since the DFT tends to overestimate the \(c'\) elastic constant. 

In the calculation of elastic constants, instead of the classical approach i.e. fitting a polynomial to an energy-strain curve and taking an appropriate derivative, we fit Brugger's equation of state (EOS) (\cite{Brugger1964}) to all relationships at the same time. We use the explicit form for cubic materials as given by Vekilov et al. in \cite{KhVekilov2016}. By doing so, we reduce the dependency of our results on the selection of deformation patterns and order of calculations. Note that, in principle, it is possible to calculate second-order elastic constants (SOEC) from at least 3 deformation patterns. There is an infinite number of patterns (although they are not arbitrary) we can use and results may vary significantly. We decided to use an equation-of-state that takes into account up to third-order constants (TOEC). For that reason, we had to limit our calculations to 0.025 strain magnitude. Otherwise, it was impossible to recreate energy-strain curves as it would require more flexible (higher-order) EOS. Also note that this meta-parameter is another way that allows us to adjust the results. Therefore, one needs to be careful when interpreting them. While SOEC are not very sensitive to such manipulations, some TOEC can be greatly affected by selected strain magnitude.

It is fairly simple to assess SOEC since high-quality experimental data are available. Here, we relied on the more recent publication by Adams et al. (\cite{Adams2006}) instead of the classical paper by Rayne (\cite{Rayne1961}). However, it is much more difficult to obtain results for TOEC at temperatures near \(0\,\text{K}\). For that reason, we used a mixture of theoretical and experimental results in the reference data set, where experimental results include also measurements at the room temperatures. Furthermore, we assess predictions by our tight-binding probabilistically. For each quantity of interest (QOI) and each model, we estimate expectation and variance. The test statistic is inspired by the \(t\text{-test}\) and will be given by
\begin{equation}
t_{\alpha\beta} = \frac{X_{\alpha\beta}-m_{\beta}}{s_{\beta}},
\end{equation}
where \(X\) is the matrix of TB results, \(\alpha\) indexes the model while \(\beta\) the QOI, \(m\) corresponds to test-set mean and \(s\) sample standard deviation. To interpret data we can compare values of \(t\) to quantiles of the normal distribution with zero expectation and unit variance -- \(\mathcal{N}(0,1)\). The most important quantiles are \(0.975/0.025\) that correspond to values \(\pm1.96\) of \(t_{\alpha\beta}\). Results are presented in Table \ref{tab:ec_fin}. 
\begin{table*}[htb!]
\small
Elastic constants in (GPa)\\
\begin{tabular}{lrrrrrrrrrrr}
\toprule
{} & \multicolumn{2}{r}{073} & \multicolumn{2}{r}{184} & \multicolumn{2}{r}{203} & \multicolumn{5}{r}{reference} \\
\cmidrule(lr){2-3}\cmidrule(lr){4-5}\cmidrule(lr){6-7}\cmidrule(lr){8-12}
{} &     val. &      t &     val. &      t &     val. &      t &         1 &      2 &        3 &        4 &        5 \\
\midrule
$c_{11}$  &   288.41 &   1.60 &   269.80 &   0.80 &   286.25 &   1.51 &    243.00 & 239.55 &   263.04 &   226.00 &   285.21 \\
$c_{12}$  &   192.57 &   4.34 &   154.31 &   0.83 &   179.69 &   3.16 &    138.00 & 135.75 &   151.14 &   140.00 &   161.61 \\
$c_{44}$  &    95.67 &  -2.56 &   104.99 &  -1.32 &    97.97 &  -2.26 &    122.00 & 120.75 &   103.34 &   116.00 &   112.51 \\
$c_{111}$ & -3038.71 &   0.95 & -4277.47 &  -0.22 & -5954.21 &  -1.81 &  -4820.00 &    - & -4971.50 & -2720.00 & -3665.98 \\
$c_{112}$ & -1195.87 & -13.45 & -1367.62 & -17.86 & -1067.33 & -10.19 &   -700.00 &    - &  -675.14 &  -608.00 &  -673.96 \\
$c_{123}$ &  -975.66 &  -0.63 & -1349.60 &  -0.86 & -2124.49 &  -1.35 &   2460.00 &    - &  -806.29 &  -578.00 &  -891.74 \\
$c_{144}$ & -1014.91 &   0.17 & -1047.92 &   0.09 &  -930.28 &   0.39 &  -1580.00 &    - &  -699.01 &  -836.00 & -1220.63 \\
$c_{155}$ &  -801.47 &   0.01 &  -557.97 &   0.90 &  -655.48 &   0.54 &  -1030.00 &    - &  -606.69 &  -530.00 & -1046.25 \\
$c_{456}$ &  -697.73 &  -0.39 &  -634.81 &  -0.27 &  -743.97 &  -0.48 &    275.00 &    - &  -647.01 &  -720.00 &  -880.67 \\
$c'$      &    47.92 &  -0.75 &    57.74 &   0.69 &    53.28 &   0.04 &     52.50 &  51.90 &    55.95 &    43.00 &    61.80 \\
$B$       &   224.52 &   2.99 &   192.81 &   0.83 &   215.21 &   2.36 &    173.00 & 170.35 &   188.44 &   168.67 &   202.81 \\
\bottomrule
\end{tabular}

\caption{Estimates of elastic constants. References are as follows: 1 -- experimental data, \cite{Hughes1953} and \cite{choy1979elastic} after \cite{pham2010lattice}, 2 -- experimental data \cite{Adams2006}, 3 -- DFT (GGA) estimate, \cite{pham2010lattice}, 4 -- experimental data, high temperatures, \cite{Blaschke2017}, 5 -- DFT (GGA) estimate, this work.}
\label{tab:ec_fin}
\end{table*}
It is clear that the model 184 provides an excellent agreement with the reference data with respect to the SOEC. However, estimates of TOEC are far from perfect. Although given the data we can test only the consistency with the set. The set itself is rather arbitrary as it is based on the availability of the data. Furthermore, uncertainty associated with data is unknown. \cmt{Squeeze comparison here with NRL (neq c12) Tonys sd and stuff?}

\section{Transferability}
\label{sec:org02e22dc}
The most important aspect of a model is its transferability. In this work, we are most concerned that our model will be able to predict energies and associated forces in bcc structures with defects. However, the predictive value can be assessed by the estimation of formation energies and elastic constants for other structures. In principle, the family of parametric tight-binding models has a significant advantage over analytical or purely numerical "surrogate" models. Although the definition of the pair-wise repulsion is arbitrary to some degree, a big part of the model is a valid approximation to the many-body problem. In this section, we investigate the behaviour of our models in non-native structures.

\begin{figure*}[htb!]

\begin{center}
\includegraphics[width=.9\linewidth]{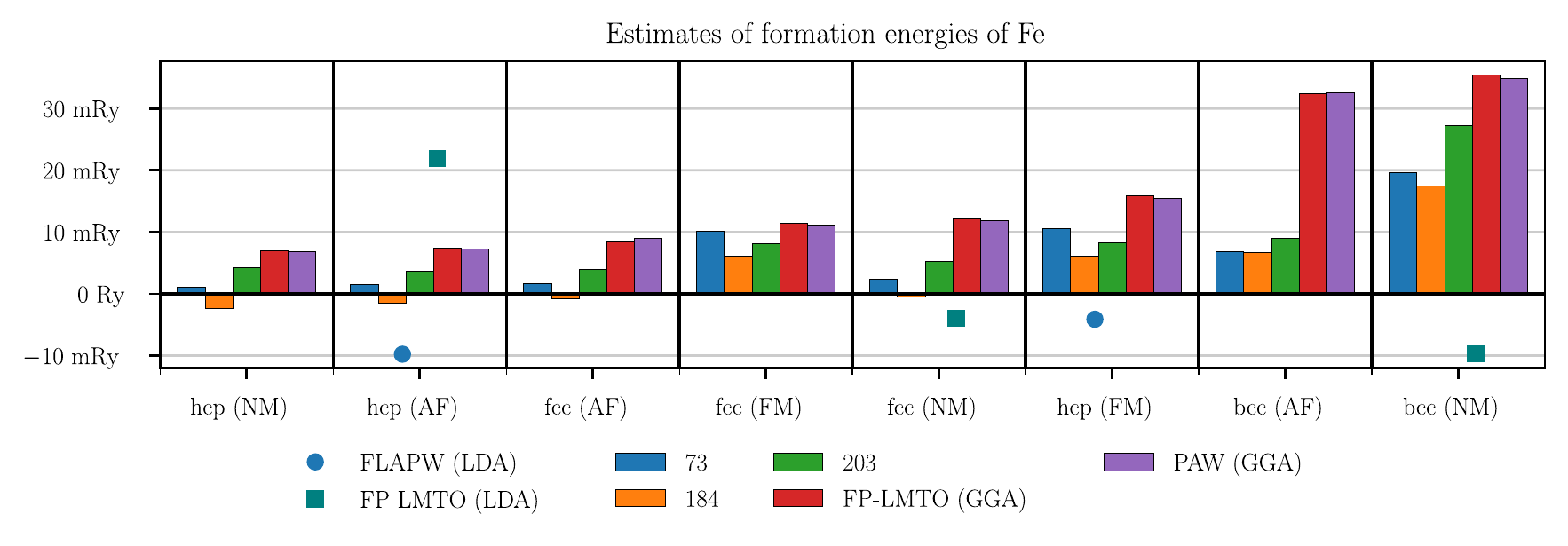}
\end{center}
\caption{Formation energies of Fe, relative to the binding energy of the bcc (FM) phase, estimated using variety of models. Results for FLAPW (LDA) are taken from \cite{Joubert1999}. LDA results are included to illustrate challenges associated with approximation of exchange and correlation effects. However, they should not be considered as the optimality limit of this method.}
\label{fig:form_en}
\end{figure*}

\begin{figure*}[htb!]
\centering
\includegraphics[width=.9\linewidth]{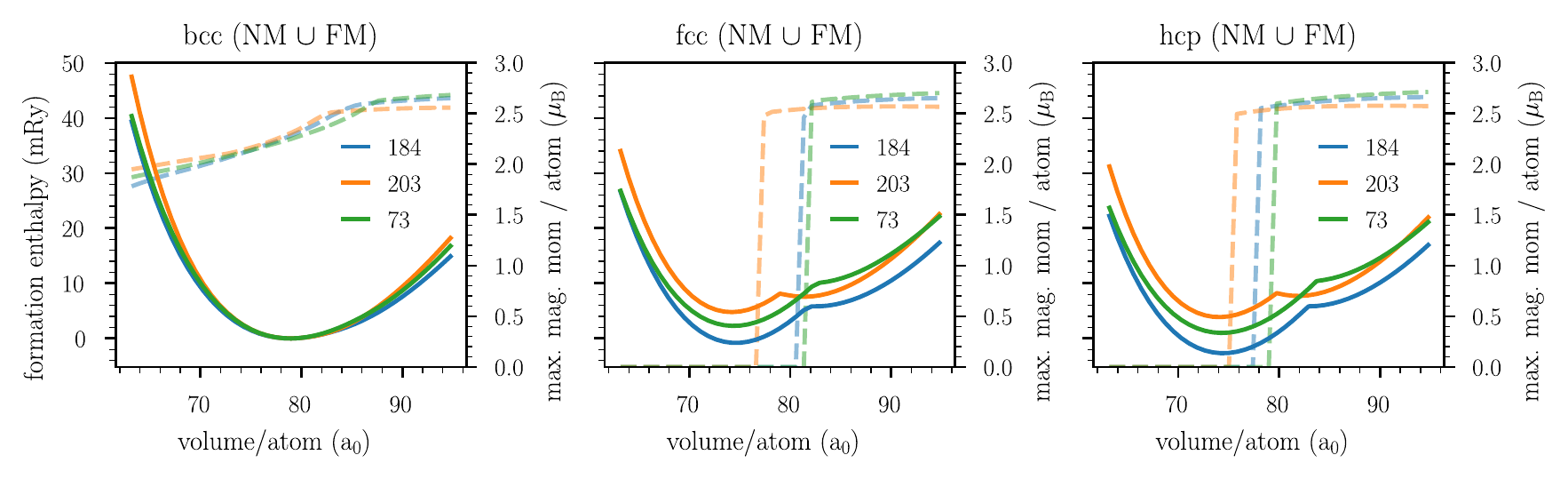}
\caption{\label{fig:org590155e}The lower value from two total energy--volume curves (FM and NM phases only), relative to the bcc formation energy. The curves represent data with piece-wise linear interpolation. The dashed lines are plotted against right axis and correspond to magnetisation per atom. Note that the the total energy is shifted by the reference energy which depends on the initial charge configuration. The purpose of these graphs is to illustrate correctly predicted transition from one magnetic state to another under the volumetric strain.}
\end{figure*}

\begin{figure*}[htbp!]
\centering
\includegraphics[width=.9\linewidth]{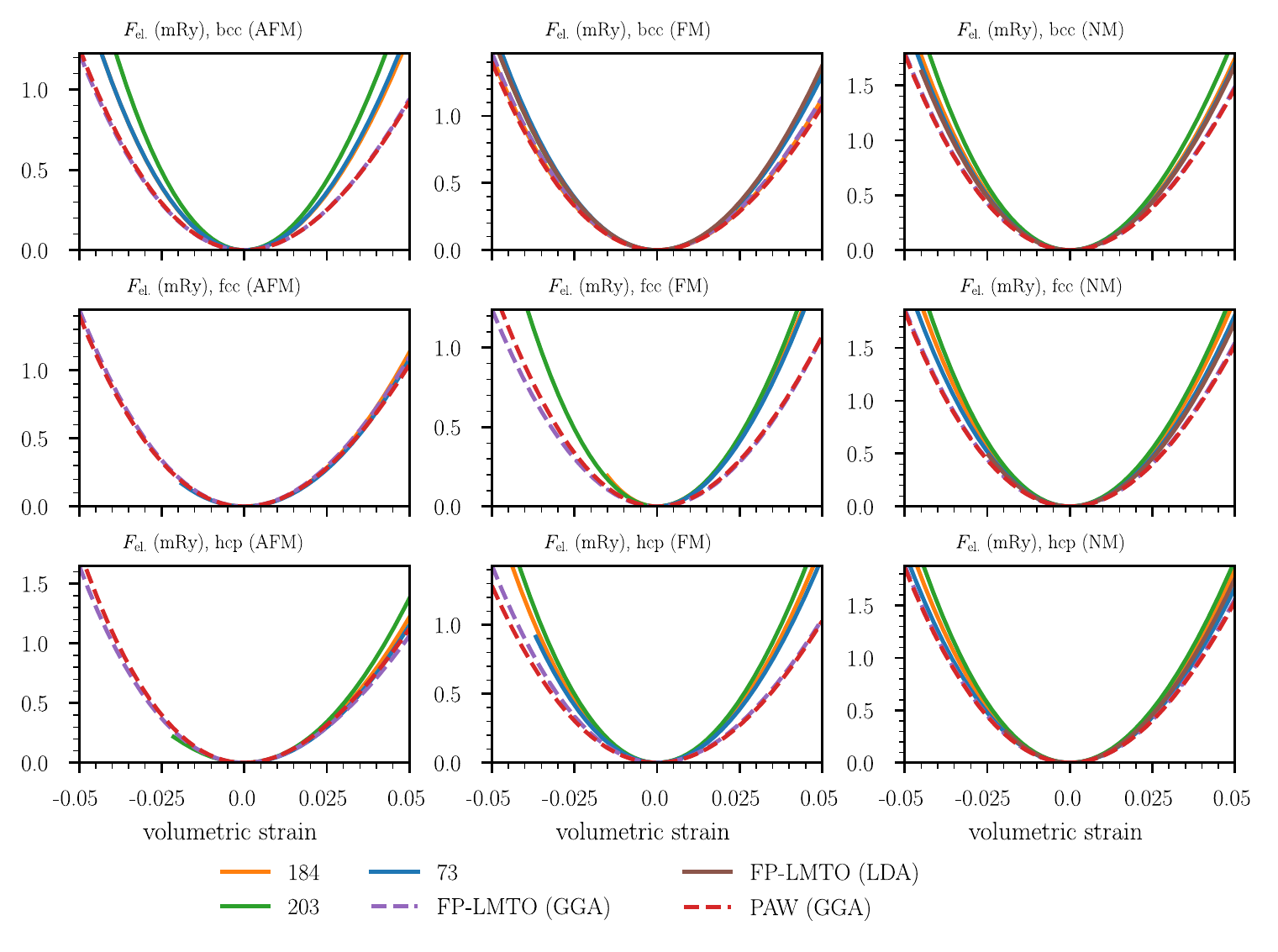}
\caption{\label{fig:orge06eb59}Estimates of elastic free energy per atom as a function of the volumetric strain in Fe. Each rows represents a given structure while column a magnetic state.}
\end{figure*}

\begin{figure*}[htb!]

\includegraphics[height=0.25\textwidth]{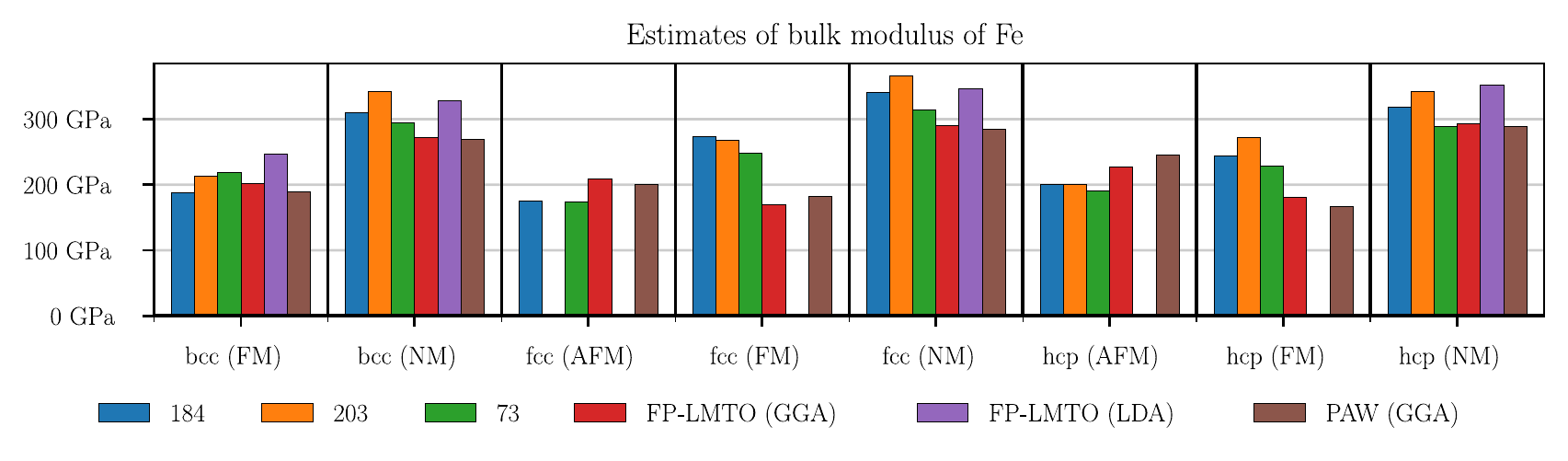}
\caption{\label{fig:org14eae2c}Estimates of the bulk modulus. In some cases, due to lack of well a defined minimum.}
\end{figure*}

The first consideration are formation energies (\(E_{\text{form.}}\)) with respect to the bcc structure in the ferromagnetic state (FM) i.e. \cmt{I need help with naming and definitions}
\begin{equation}
\label{eq:orgb153fcc}
E_{\text{form.}}\left(\text{struc.}\right)=\min_{V}\left\{ E\left(V|\text{struc.}\right)-E\left(V_{0}|\text{bcc}\right)\right\},
\end{equation}
where the function \(\text{min}\) returns the minimum of a set, and \(\text{struc.}\) refers to the crystallographic structure in question. Minimisation is carried out over volume per atom \(V\) and \(V_0\) corresponds to the ground state of the bcc (FM) structure at \(0\,\text{K}\). In practice, we evaluated energies over the range from -0.2 to 0.2 of engineering volumetric strains with a 0.005 increment. Note that not all calculations completed successfully. For example, calculations with the ferromagnetic starting point were likely not to converge near strains that favour the non-magnetic state. Results were interpolated by fitting a 12th order polynomial
\begin{equation}
\label{eq:org8d8195d}
E(V|\text{struc.}) = \sum_{n=0}^{12} \alpha_{n} \left(\frac{1}{V^{1/3}} \right)^n.
\end{equation}
In the process we used the standard least-square methods. The form of the function was inspired by the SJEOS equation of state \cite{Alchagirov2001} with some extra degrees of freedom added to improve the fit. In this case we don't require high numerical precision and any reasonable interpolator will suffice. The focus is on bcc, fcc and hcp structures in ferromagnetic, non-magnetic (NM) and anti-ferromagnetic (AFM) phases. We consider non-co-linear magnetism with a single-layer on an arbitrary closed-packed direction in the AFM case. Such spin configuration does not minimise the total energy although should be sufficient for comparison. We begin with analysis of a wide range of methods. The data presented in Figure \ref{fig:form_en} consists of the following: results from TB calculations -- model 73, 184, 203; full-potential LMTO (linearised muffin-tin orbitals) method implemented in Questaal (\cite{Pashov2019}) with GGA (PBE) and LDA functionals (FP-LMTO (GGA) and FP-LMTO (GGA) respectively); PAW (projector-augmented-wave) method with GGA implemented in VASP \cite{Kresse1996}.
In summary, we have the following categories of results: TB, DFT with different exchange-correlation functionals and DFT with two types of basis-set. Before addressing TB results we consider DFT results first. Calculations in Questaal and VASP were made by the present authors and these should be considered as the main point of reference. The data from the literature shows results from the development of pseudo-potentials. The main reason for including them, as well as results from LDA calculations, is the management of expectations and validation of primary references. It immediately transpires that Questaal and VASP results are almost identical and the basis set, if optimised, does not influence results. Furthermore, the LDA results show negative formation energies for non-magnetic close-packed structures. This is the most important result in the context of what can be expected from TB models. In other words, we cannot expect exact estimates from TB when ab-initio methods are predicting different than bcc (FM) ground states of Fe at \(0\,\text{K}\). This is indeed the case in LDA calculations. From comparison with the literature, we also know that is not a result of an error on our part. Figure \ref{fig:form_en} provides quantitative assessment of TB predictions. As expected, we can hope for qualitative agreement at best and only with respect to the ordering of non-equilibrium phases. The only exception here is the fcc (FM) case, where formation energy is greater than the overall trend would suggest. However, note that the initial charge distribution affects reference energy in TB calculations. This makes it difficult to assess the results. Nonetheless, the results are directly comparable within a given magnetic phase. Next, we consider the transition from the magnetic state under pressure. If magnetism is correctly included in the model, such transformation should occur in closed-packed structures. Predictions of the TB models are illustrated in Figure \ref{fig:org590155e}.
In a broader sense, our models behave as they should. This is consistent with DFT results. Namely, within the considered range of strains, the FM state stabilises the bcc structure. Close-packed structures will experience a transition from the FM to NM state under compression. All models predict similar transitions points which is consistent with DFT results. The above also illustrates that in some cases there is no minimum in the metastable ferromagnetic state. The estimations of formation energies were based on extrapolation of a 3rd-order polynomial fitted to several last available data-points. In the context of transferability into structures with defects, mechanical properties in non-equilibrium structures might provide a better indication of performance. We limit our calculations to estimates of elastic free energy and calculations of the bulk modulus. While this is not a sufficient indicator it is a necessary one. The results are presented in Figure \ref{fig:orge06eb59}.
The performance varies significantly depending on the structure and magnetic phase. Differences in estimates for the bcc (FM) case are quite small as this was included in the fitness function. We can see that TB and DFT (including GGA and LDA) are in a very good agreement. Furthermore, we can see a fair agreement in non-magnetic phases and mostly very good agreement in anti-ferromagnetic phases. We emphasise that Figure \ref{fig:orge06eb59}  shows elastic free energy per atom versus volumetric strain. Estimates of elastic energy density would be different since each method predicts different equilibrium volume. While TB was forced to fit the experimental lattice parameter (\(2.866\,\text{\AA}\)) of the bcc (FM) structure, DFT with GGA functional would slightly underestimate it. Predictions of the DFT method within the LDA resulted in significant underestimations. In some cases, this method failed to find a minimum even within the range of \(\pm0.2\) strain (with the experimental lattice parameter for bcc Fe as a reference point). This reflects on better estimates of bulk modulus in case of the TB method and much worse performance of the LDA than Figure \ref{fig:orge06eb59} would suggest. This is clearly illustrated in Table \ref{tab:bulk_val} and Figure \ref{fig:org14eae2c}.
\begin{table*}[htb!]
\small
\begin{tabular}{lrrrrrrrrr}
\toprule
{} &  bcc (AFM) &  bcc (FM) &  bcc (NM) &  fcc (AFM) &  fcc (FM) &  fcc (NM) &  hcp (AFM) &  hcp (FM) &  hcp (NM) \\
\midrule
184           &      225.0 &     187.9 &     309.5 &      175.7 &     273.4 &     341.5 &      200.9 &     243.4 &     318.0 \\
203           &      278.0 &     213.0 &     342.5 &         -  &     268.3 &     366.8 &      200.2 &     272.7 &     342.4 \\
73            &      225.9 &     219.1 &     294.4 &      174.1 &     248.2 &     313.6 &      190.2 &     228.9 &     288.5 \\
FP-LMTO (GGA) &      172.7 &     201.8 &     272.6 &      209.0 &     169.5 &     289.7 &      227.2 &     180.8 &     293.7 \\
FP-LMTO (LDA) &         -  &     247.4 &     328.1 &         -  &        -  &     346.3 &         -  &        -  &     351.5 \\
PAW (GGA)     &      174.2 &     189.5 &     269.4 &      200.3 &     182.6 &     285.4 &      244.8 &     166.7 &     289.2 \\
\bottomrule
\end{tabular}

\caption{Estimates of bulk modulus of Fe at $0\,\mathrm{K}$.}
\label{tab:bulk_val}
\end{table*}

In summary, the performance of the TB method in this test is no worse than LDA. Except for the bcc (FM) case, the earliest model no. 73 seems to perform better than its counterparts. We can also see that these estimates differ from ones presented in Table \ref{tab:ec_fin}. In this test, we simply report the normalised value of the second derivative at equilibrium instead of fitting a complete model of the elastic energy as a function of the strain. Therefore, a slightly worse performance of the model 203 needs to be considered with some extra care. Nonetheless, we focus here on the large deviations from the reference. We assume that the problem arises from magnetic interactions for phases that are far from the equilibrium. For example, a test on fcc (AFM) demonstrated a fairly good agreement with DFT (GGA) calculations. On the other hand, fcc (FM) performed quite poorly. According to predictions by Wr\'{o}bel et al. (\cite{Wrobel2015}) the ground state of fcc phase (at \(0\text{K}\)) corresponds to the anti-ferromagnetic double-layer (AFMDL) spin configuration. It seems that the pattern is as follows: the further away we are from the ground state -- in terms of magnetic configuration, the less consistent with the DFT (GGA) are the estimates of elastic free energy. This effect includes calculations for fcc and hcp structures. Although we can only speculate at this point, these results have a reasonable explanation. The non-selfconsistent Hamiltonian (the starting point of the TB calculations) at best corresponds to a variational approximation to the ground state density. Implicitly, by setting parameters, we specify a starting point on which a variational approximation depends. In such context, it seems that there is a limited energy window in which TB models can be transferable. On the other hand, it is also possible that by prioritising a good fit of the band structures we underestimated the overlaps and lost transferability of the hopping integrals through inaccurate consideration of the environmental dependence (see e.g. Paxton et al. in \cite{Paxton2010} and \cite{Paxton2013}).

\section{Defects}
\label{sec:orgfd1593d}
\begin{figure}[htbp!]
\centering
\includegraphics[height=0.4\textwidth]{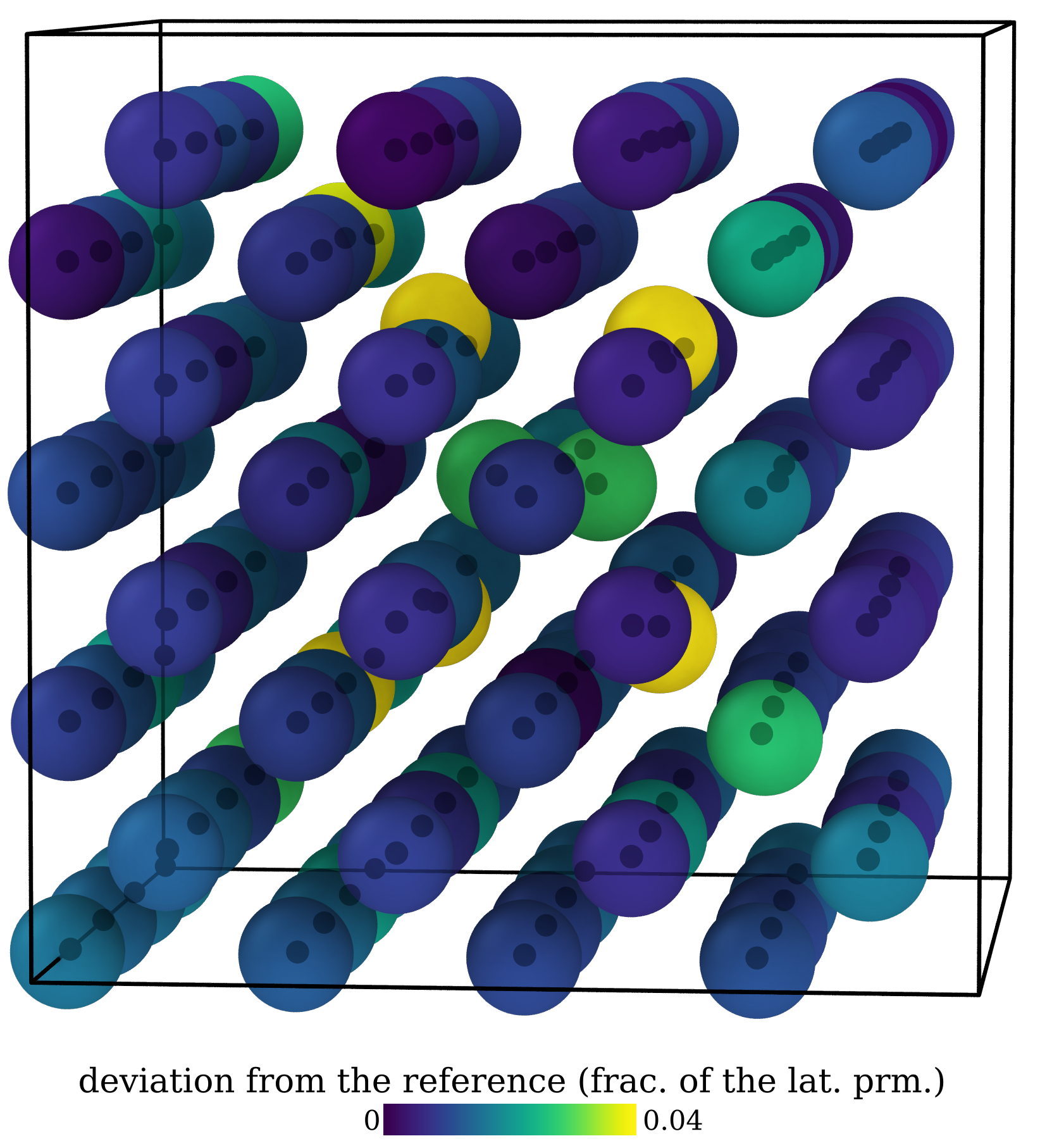}
\caption{\label{fig:org1d60f7b}Difference between dumbell structures as estimated by the DFT and TB (model 184). The difference is given in fractions of the equilibrium lattice parameter for the bcc Fe (FM) as predicted by the given method. Black circles are representing DFT estimates and spheres the TB. Colour corresponds to the norm of differences between relative positions. In case of the vacancy, structure deviations are no greater than 2\% of the corresponding lattice parameter. The acronym GAP corresponds to Gaussian approximation potential that uses machine-learning methods to estimate an optimal classical potential surrogate \cite{Bartok2010}.}
\end{figure}

In the scope of this paper, the most important tests of transferability are the self-interstitial and mono-vacancy formation energies (\(E_\text{i(v)}^\text{f}\)). This relatively small test set will provide a measure of the predictive value of the model. These energies are given by
\begin{equation}
E_\text{i(v)}^\text{f}= E_\text{i(v)} -  \left( \frac{N \pm 1}{N}\right) E_\text{bulk},
\end{equation}
where \(E_\text{bulk}\) is the total energy of a computational cell, containing \(N\) atoms, without the defect, while \(E_\text{i}\) and \(E_\text{v}\) are the total energy of a cell containing a self-interstitial or a vacancy respectively. Estimates are made for a variety of volumes with fully relaxed atomic positions. The results are presented in Figure \ref{fig:def_form_en} and summarised in Table \ref{tab:orge9b1455}. Figure \ref{fig:org1d60f7b} shows comparison of the interstitial structure as predicted by DFT (PAW with GGA) and TB. The difference will be given by the Euclidean (\(L_2\)) norm
\begin{equation}
\left\Vert \vec{x}_\text{DFT}/a_{0\text{(DFT)}} - \vec{x}_\text{TB}/a_{0\text{(TB)}} \right\Vert_2,
\end{equation}
where \(a_{0\text{(DFT/TB)}}\) is the bcc lattice parameter predicted by the corresponding method, while \(\vec{x}_\text{DFT/TB}\) is the position of an atom.
\begin{table}[htb!]
\caption{\label{tab:orge9b1455}Defect formation energies, given in electronvolts, where \(\Delta E^\text{f}\) is the difference between estimates for the \(\left<110\right>\) dumbbell and a monovacancy. The following references were used: ()\ts{1} -- \cite{Dragoni2018}, ()\ts{2} -- \cite{Liu2005}, ()\ts{3} and ()\ts{4} -- multiple sources in \cite{Dragoni2018}. \cmt{Being lazy, is this acceptable?}}
\centering
\small
\begin{tabular}{lrrr}
\hline
model & \(E_\text{v}^\text{f}\) & \(E_{\text{i}\,\left<110\right>}^\text{f}\) & \(\Delta E^\text{f}\)\\
\hline
73 & 1.87 & 2.72 & 0.85\\
184 & 2.60 & 3.60 & 1.00\\
203 & 2.03 & 2.85 & 0.82\\
GAP\ts{1} & 2.26 & 4.21 & 1.95\\
TB (d)\ts{2} & 1.75 & 4.36 & 2.61\\
DFT\ts{3} & 2.07-2.15 & 3.64-4.02 & -\\
experiment\ts{4} & 0.55-2.00 & 4.70-5.00 & -\\
\hline
\end{tabular}
\end{table}

\begin{figure}[htbp!]
\centering
\includegraphics[height=0.3\textwidth]{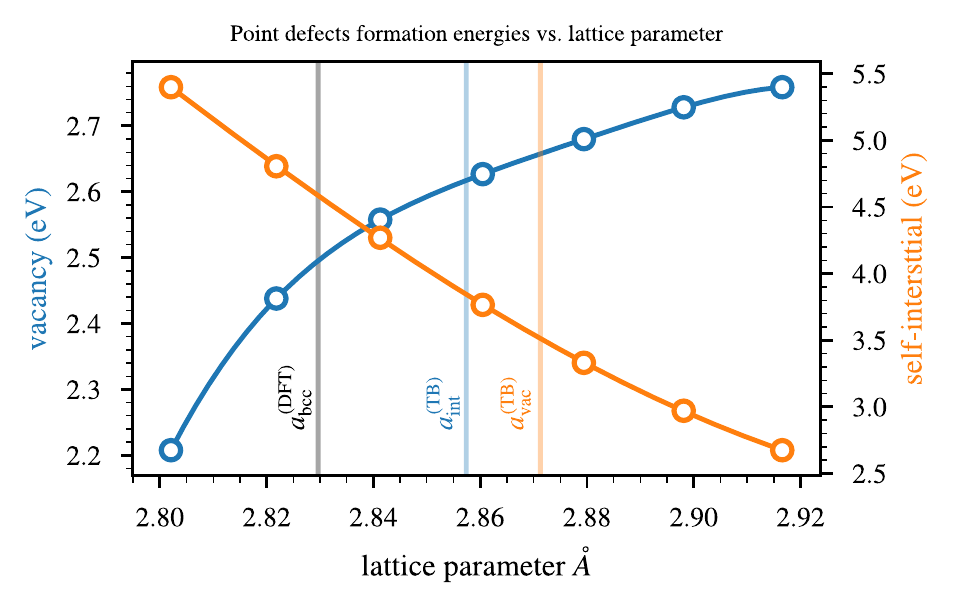}
\caption{\label{fig:org4b8b638}Comparison of formation energies \cmt{enthalpies?} of a vacancy and self-interstitial (model 184) as a function of the lattice parameter.}
\end{figure}
\begin{figure*}[htb!]
\begin{center}
\includegraphics[height=0.25\textwidth]{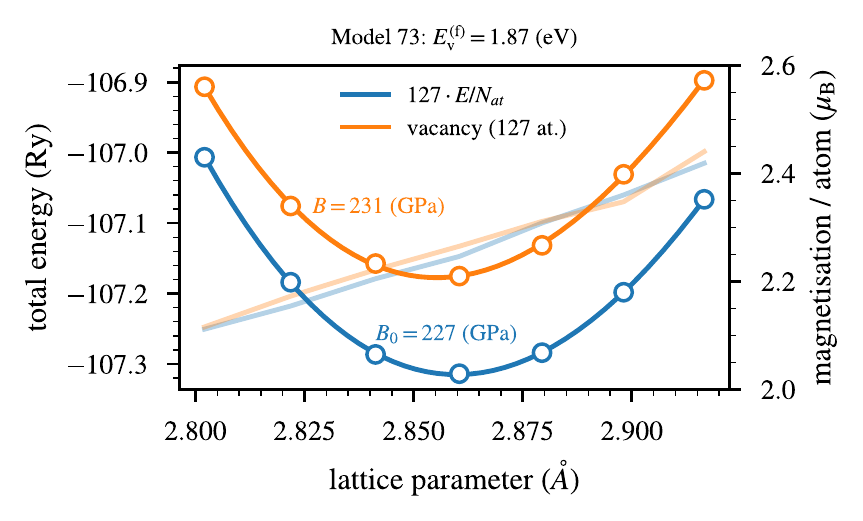}  \includegraphics[height=0.25\textwidth]{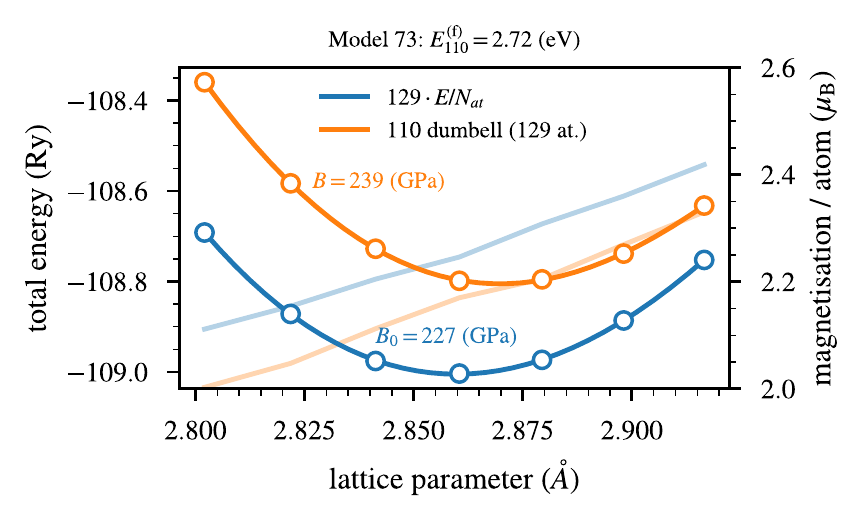}
\includegraphics[height=0.25\textwidth]{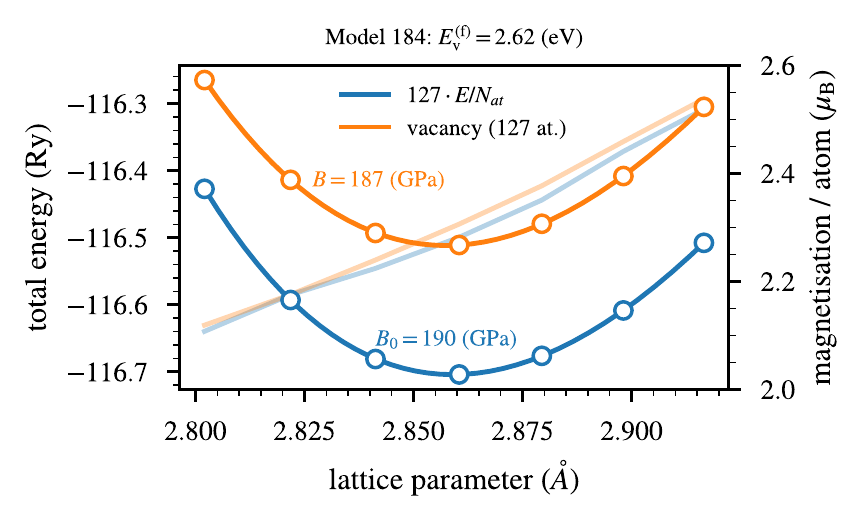} \includegraphics[height=0.25\textwidth]{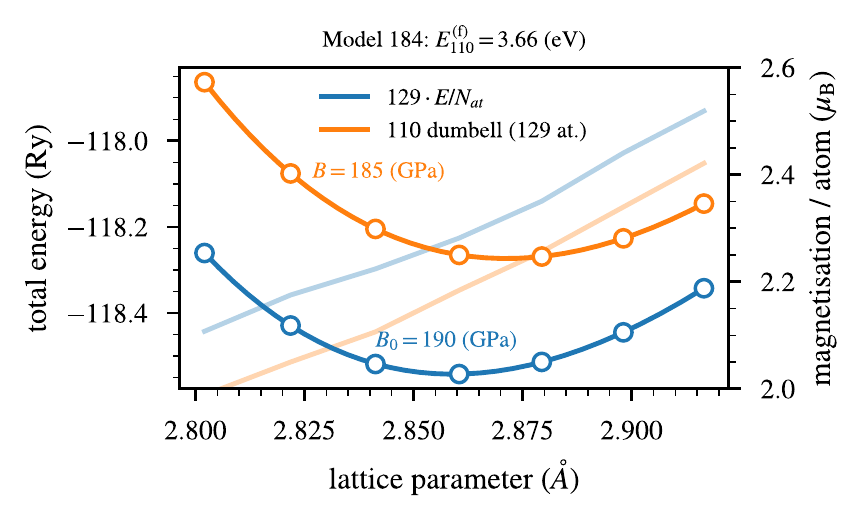}
\includegraphics[height=0.25\textwidth]{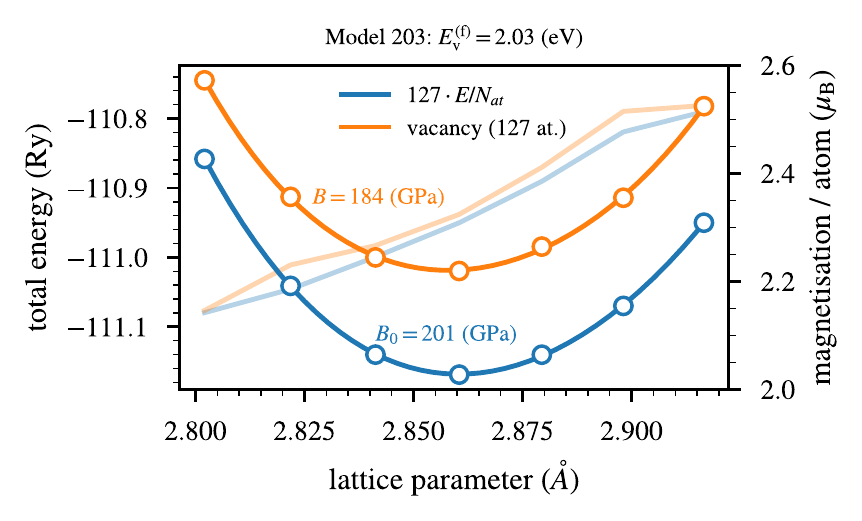} \includegraphics[height=0.25\textwidth]{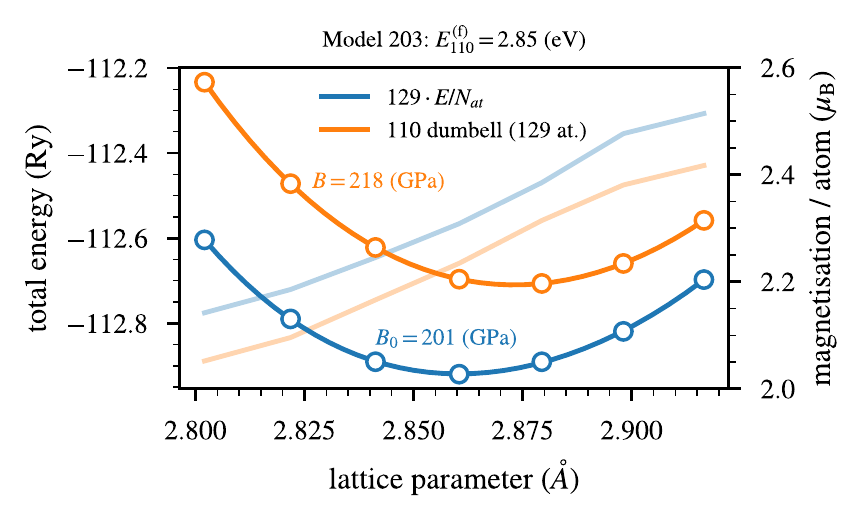}
\end{center}
\caption{Formation energies of defects in the pure bcc (FM) iron. The bulk modulus for the reference energy-volume curve (blue lines) is calculated before scaling. In calculations, we used 4th order polynomial and evaluated the value of the 2nd derivative at the minimum. Calculations were made with the initial $4 \times 4 \times 4$ number of $k\text{-points}$ in a $(4 \times 4 \times 4) \times a_\text{bcc}^3$ computational cell. Such sampling usually results in an extra 2\% numerical uncertainty associated with estimates of the formation energies.}
\label{fig:def_form_en}
\end{figure*}
We begin with the consideration of three major characteristics of the energy-volume curves (Figure \ref{fig:def_form_en}): lattice parameter, magnetisation and bulk modulus. Unlike the self-interstitial, introduction of a vacancy has no major effect on the first two quantities. As one might expect, the introduction of an extra atom reduces the magnetisation per atom and increases the equilibrium lattice parameter. Values of the bulk modulus decreased (vacancy) or remained largely unchanged (interstitial). Qualitatively, these changes can be considered as reasonable. However, estimates of elastic constants are extremely sensitive. Most likely this is a numerical artefact arising from the insufficient number of \(k\text{-points}\). It would appear that models 73 and 203 provided a very good estimate of the vacancy formation energy while failing in the case of an interstitial defect. On the other hand, the difference between the lowest DFT estimate of the interstitial formation energy and worst TB prediction (model 73) is below 25\%. Model 184 is consistent with the DFT predictions of interstitial formation energy while giving a bit too high estimates in case of a vacancy (at least 30\% higher than DFT). Nonetheless, differences between formation energies suggest that this model is performing better. So far it seems that by targeting elastic constants we can improve our predictions of the defect formation energies, although other factors might be important as well. For example, TB models were optimised towards \(2.2\, \mu_\text{B}\) magnetisation per atom in the perfect bcc cell. It is possible that this resulted in the underestimation of the magnetic contributions and we should sacrifice this property. However, in the optimisation of the elastic constants, higher values of the Stoner parameter, that controls the magnetisation, resulted in undesirable changes in energy-strain curves. It is also possible that we need to manage our expectations and our estimates are as good as they can get without over-fitting (sacrificing predictive value for the sake of improvement of known quantities) or the introduction of cancelling errors. In terms of structures of the defects, we obtained a very good agreement with DFT results, as illustrated in Figure \ref{fig:org1d60f7b}. Furthermore, when one considers formation energies for a fixed lattice parameter (Figure \ref{fig:org4b8b638}), it is apparent that results are more consistent for bond lengths predicted by the DFT method. Finally, in the case of a vacancy defect, we are quite close to GAP calculations that contained a vacancy in the fitting set. These findings are likely to be relevant, given that mechanical properties are based on energy-strain curves calculated using this method. In other words, we might need a better than DFT method to obtain a reference for optimisation of the mechanical properties. \cmt{Discussion of Fig. 11 should be improved.}
\section{Summary and conclusions}
\label{sec:org64107a1}
In this paper, we investigated a method of obtaining parameters for the self-consistent polarisable tight-binding model of Fe, to study diffusion in low-alloy steels in the future. We decided to use the non-orthogonal \(spd\) basis which means that \(3d\), \(4s\) and \(4p\) atomic orbitals are implicitly included in the construction of the initial Hamiltonian with the Slater-Koster algorithm. As a result, it was possible to recreate all features of the band plots for bcc, fcc and hcp structures, giving this method a potential to be fully automated. Additionally, as band calculations are fairly inexpensive, it was possible to optimise the model with a minimum number of constraints. This led to another advantage of our approach i.e. the ability to include in the objective data from methods more "complete" than the DTF, such as QSGW. Furthermore, we introduced a more robust optimisation of pair-potentials that takes into account higher-order terms. As demonstrated in Figure \ref{fig:org3866ec6}, estimation of second-order elastic constants is insufficient to confirm the correctness of the pair-potentials. This way we achieved a better sampling of pair-potentials that is more sensitive to pairs outside the 1st neighbour shell. Finally, we were able to address some interesting aspects of how to design an optimisation strategy. This includes a selection of the fitness function, optimisation algorithms and a strategy. We found that the covariance matrix adaptation strategy (CMAES) is a good algorithm to optimise a large set of tight-binding parameters, while the simplicial homology global optimisation (SHGO) is superior in the optimisation of pair-potentials. Note that the SHGO algorithm is designed for a small number of dimensions (<10), hence it couldn't be used in the optimisation of band structures. The novelty of our approach can be viewed as addressing the problem as a maximisation of the predictive value rather than minimisation of the fitness function over a rich test-set. This means we decided to optimise parameters on a small, computationally inexpensive set while iteratively focusing the search region near the most promising candidates, evaluated in subsequent stages.

Our approach was tested in a series of numerical tests aimed at a recreation of key properties of allotropes of Fe. We were able to achieve a very good agreement with reference band plots, generated using the QSGW method. It is likely that this was possible thanks to a limited number of constraints and the full \(spd\) basis. Surprisingly, out of three tested references (DFT (LDA), DFT (GGA) and QSGW), QSGW-based band structures also resulted in the best representation of the elastic properties. \cmt{Is it useful somehow?} Although we achieved a good agreement with experimental second-order elastic constants, representation of third-order elastic constants can be regarded as qualitative at best. In the context of transferability, the results are consistent with DFT (GGA) results assuming that the formation energy of a given phase does not deviate too much from the bcc (FM) binding energy. While energy-volume curves for non-magnetic phases are in a good agreement with the reference, clearly results associated with non-native ferromagnetic phases could be improved. The problem can be traced to putting too much emphasis on the representation of the band structures. It is possible that we explored a basin of orthogonal models that can have a limited transferability. Figure \ref{fig:orge1ba165} illustrates how difficult is to separate both cases. The data suggest that magnetism could be responsible for the under-performance and adjustments of the Stoner and Hubbard-like \(U\) parameter, might improve the results. On the other hand, the performance of our tight-binding model is comparable to other models. For example, Bacalis et al. underestimated the bulk modulus of the ferromagnetic bcc Fe (\cite{Bacalis2001}) as did Madsen et al. (\cite{Madsen2011}). However, in the latter paper, we can find very good results with respect to the bulk modulus in other phases. Liu et al. (\cite{Liu2005}) achieved overall very good values of bulk modulus, except for ferromagnetic and anti-ferromagnetic hcp phase. Additionally, if we consider the model by Paxton et al. \cite{Paxton2010}, and also consider estimates of bulk modulus as a simplified measure of transferability, it seems like this aspect is universally problematic. \cmt{Do we need a table? Put the damn table!}. It is needless to say, that a single elastic constant is not a good performance measure although deviations from the reference value suffice to indicate level of the transferability. In the context of our research, the most important test was the estimation of formation energies of the self-interstitial and the vacancy. Predicted structures of point defects are in good agreement with the DFT results. While it is a less robust test than e.g. evaluation of a dislocation structure, it illustrates that a model (as a set of parameters) generated using our method can be used as a low-fidelity model to speed-up relaxation of atomic positions while assisting DFT calculations. Given the influence of magnetism, this result can be considered as a positive indication. \cmt{Otherwise any pp should be able to do it.} With respect to formation energies, when comparing our results with the whole range of those generated with the DFT method, the performance of our models could be viewed as satisfactory. However, it is far from the experimental values. Furthermore, the d-band TB model by Liu et al. (\cite{Liu2005}) performed much better as well as the GAP developed by Dragoni et al. (\cite{Dragoni2018}) even though magnetism cannot be treated appropriately with purely empirical models. In defence of our model, we can point out that these results were not part of the fitting data-set and should be regarded as a pure prediction. Therefore, 20\% -- 30\% deviations from DFT predictions is something that one might expect. Having said that, the utility that our model can provide might be limited. The solute–vacancy binding energies in iron-based alloys are usually of the order of magnitude of \(0.1\,\text{eV}\) (see e.g. Messina et al. \cite{Messina2013} or Whiting et al. \cite{Whiting2019}). Although, in the context of potential applications of the TB method, more interesting are estimates of potential barrier in simulations of defect migration. In this associated errors might be at the same order of magnitude as quantities of interest. We speculate that this problem is partially associated with forcing TB models to stabilise the equilibrium lattice parameter (Fig. \ref{fig:org4b8b638}) (as well as with difficulty of optimising overlaps together with hopping integrals). Self-consistent tight-binding is consistent with the local spin-density approximation which tends to heavily underestimate the bond length of Fe. By including \(s\) and \(p\) orbitals we introduce an extra level of ambiguity as well as add interactions that cannot be well represented in this method (e.g. hopping between \(p\) and \(d\) atomic orbitals). \cmt{Maybe overlap is too big there?} 

Despite the issues  encountered we still think that our approach is an interesting path to explore and multiple improvements can be introduced. For example, currently, we are investigating the possibility of improving the optimisation procedure by using all data generated so far. We aim to find an appropriate basis to describe the essential features of the band structures so we can find an inverse of the fitness function. This would allow us to investigate which parameters can be changed without sacrificing the fit to the bands and test parameters from this subset to find ones that satisfy mechanical properties. As a result, we would be also able to provide some approximation of the uncertainty of the band structure estimation. \cmt{Kernel regression?} Further improvement could involve testing a bigger variety of functions describing decays and the pair-potential, including cut-off radii in the optimisation stage or select functions that involve screening effects. There is also the somewhat controversial option of using machine-learning methods to find an exact interpolation of the optimal pair-potential, which might, but need not be, environmentally dependent. On top of the optimisation methodology, we also consider improvements in the reference data-set that could include calculation of some properties associated only with the electronic structure, such as magnetic susceptibility etc.
\cmt{With large enough data-st we can sort out constraints How's that useful? Maybe we can sort-out the overlap with some transf. tests. Proper treatment like ML shoud solve the issue. The question is, if it's worth it?}
\section{Acknowledgments}
\label{sec:org37f3c07}
We would like to kindly acknowledge The Engineering and Physical Sciences Research Council (EPSRC) for funding this project (ref. EP/P003591/1). We would like to also thank Dr Dimitar Pashov and Dr Mark Wenman for their valuable comments and continuous support. CP Race was founded by a University Research Fellowship of the Royal Society. Calculations were performed on a computational cluster maintained by Computational Shared Facility of the University of Manchester.
\cmt{Who (what) else?}
\newpage

\section{References}
\label{sec:orgdea055c}

\bibliographystyle{unsrt}
\bibliography{bib}

\begin{thebibliography}{10}

\bibitem{Slater1954}
J~C Slater and G.~F. Koster.
\newblock {Simplified LCAO method for the periodic potential problem}.
\newblock {\em Physical Review}, 94(6):1498--1524, 1954.

\bibitem{Andersen1984}
O.~K. Andersen and O.~Jepsen.
\newblock {Explicit, first-principles tight-binding theory}.
\newblock {\em Physical Review Letters}, 53(27):2571--2574, dec 1984.

\bibitem{Cohen1994}
Ronald~E. Cohen, Michael~J. Mehl, and Dimitrios~A. Papaconstantopoulos.
\newblock {Tight-binding total-energy method for transition and noble metals}.
\newblock {\em Physical Review B}, 50(19):14694--14697, 1994.

\bibitem{Mehl1996}
Michael~J Mehl and Dimitrios~A Papaconstantopoulos.
\newblock {Applications of a tight-binding total-energy method for transition
  and noble metals: Elastic constants, vacancies, and surfaces of monatomic
  metals}.
\newblock {\em Physical Review B - Condensed Matter and Materials Physics},
  54(7):4519--4530, 1996.

\bibitem{Kohn1965}
W~Kohn and L~J Sham.
\newblock {Self-consistent equations including exchange and correlation
  effects}.
\newblock {\em Physical Review}, 140(4A), 1965.

\bibitem{Paxton2008}
Anthony~T. Paxton and Michael~W. Finnis.
\newblock {Magnetic tight binding and the iron-chromium enthalpy anomaly}.
\newblock {\em Physical Review B - Condensed Matter and Materials Physics},
  77(2), 2008.

\bibitem{Stoner1938}
C~Stoner.
\newblock {Collective electron ferronmagnetism}.
\newblock {\em Proceedings of the Royal Society of London. Series A.
  Mathematical and Physical Sciences}, 165(922):372--414, 1938.

\bibitem{Bacalis2001}
N.~C. Bacalis, D.~A. Papaconstantopoulos, M.~J. Mehl, and M.~Lach-Hab.
\newblock {Transferable tight-binding parameters for ferromagnetic and
  paramagnetic iron}.
\newblock {\em Physica B: Condensed Matter}, 296(1-3):125--128, feb 2001.

\bibitem{Liu2005}
Guoqiang Liu, D~Nguyen-Manh, Bang~Gui Liu, and D~G Pettifor.
\newblock {Magnetic properties of point defects in iron within the
  tight-binding-bond Stoner model}.
\newblock {\em Physical Review B - Condensed Matter and Materials Physics},
  71(17), 2005.

\bibitem{Paxton2010}
A~T Paxton and C~Els{\"{a}}sser.
\newblock {Electronic structure and total energy of interstitial hydrogen in
  iron: Tight-binding models}.
\newblock {\em Physical Review B - Condensed Matter and Materials Physics},
  82(23), 2010.

\bibitem{Madsen2011}
Georg~K.H. Madsen, Eunan~J. McEniry, and Ralf Drautz.
\newblock {Optimized orthogonal tight-binding basis: Application to iron}.
\newblock {\em Physical Review B - Condensed Matter and Materials Physics},
  83(18):184119, 2011.

\bibitem{Hatcher2012}
Nicholas Hatcher, Georg K~H Madsen, and Ralf Drautz.
\newblock {DFT-based tight-binding modeling of iron-carbon}.
\newblock {\em Physical Review B - Condensed Matter and Materials Physics},
  86(15):155115, 2012.

\bibitem{Horsfield1998}
Andrew Horsfield and Steven~David Kenny.
\newblock {Efficient ab initio tight binding}.
\newblock In {\em Materials Research Society Symposium - Proceedings}, volume
  491, pages 57--63, 1998.

\bibitem{Kotani2007}
Takao Kotani, Mark {Van Schilfgaarde}, and Sergey~V Faleev.
\newblock {Quasiparticle self-consistent GW method: A basis for the
  independent-particle approximation}.
\newblock {\em Physical Review B - Condensed Matter and Materials Physics},
  76(16), 2007.

\bibitem{gw_tutorial}
See {QSGW} tutorial for magnetic bcc {F}e:
  \url{https://www.questaal.org/tutorial/gw/qsgw_fe/}.

\bibitem{Paxton2009a}
Anthony~T. Paxton.
\newblock {An Introduction to the Tight Binding Approximation –
  Implementation by Diagonalisation}.
\newblock In Dominik~Marx {Johannes Grotendorst, Norbert Attig, Stefan
  Bl{\"{u}}gel}, editor, {\em Multiscale Simulation Methods in Molecular
  Sciences}, volume~42, pages 145--176. Forschungszentrum J{\"{u}}lich,
  J{\"{u}}lich, 2009.

\bibitem{sutton1997interfaces}
A~P Sutton and R~W Balluffi.
\newblock {\em {Interfaces in Crystalline Materials}}.
\newblock 1997.

\bibitem{Finnis2010}
Mike Finnis.
\newblock {\em {Interatomic Forces in Condensed Matter}}.
\newblock Oxford University Press, jan 2010.

\bibitem{Finnis1997}
M~{Finnis, M. W. and Paxton, A. T. and Methfesselt, M. and van Schilfgaarde}.
\newblock {Self-Consistent Tight-Binding Approximation Including Polarisable
  Ions}.
\newblock {\em Tight-Binding Approach to Computational Materials Science},
  pages 265--274, 1997.

\bibitem{Sutton1988}
A.~P. Sutton, M.~W. Finnis, D.~G. Pettifor, and Y.~Ohta.
\newblock {The tight-binding bond model}.
\newblock {\em Journal of Physics C: Solid State Physics}, 21(1):35--66, jan
  1988.

\bibitem{Pashov2019}
Dimitar Pashov, Swagata Acharya, Walter R.~L. Lambrecht, Jerome Jackson,
  Kirill~D. Belashchenko, Athanasios Chantis, Francois Jamet, and Mark van
  Schilfgaarde.
\newblock {Questaal: a package of electronic structure methods based on the
  linear muffin-tin orbital technique}.
\newblock {\em Computer Physics Communications}, 2019.

\bibitem{PhysRevLett.81.5149}
M.~W. Finnis, A.~T. Paxton, M.~Methfessel, and M.~van Schilfgaarde.
\newblock Crystal structures of zirconia from first principles and
  self-consistent tight binding.
\newblock {\em Phys. Rev. Lett.}, 81:5149--5152, Dec 1998.

\bibitem{Hubbard1963}
J.~Hubbard.
\newblock {Electron correlations in narrow energy bands}.
\newblock {\em Proceedings of the Royal Society of London. Series A.
  Mathematical and Physical Sciences}, 276(1365):238--257, nov 1963.

\bibitem{Harrison1985}
Walter~A Harrison.
\newblock {Coulomb interactions in semiconductors and insulators}.
\newblock {\em Physical Review B}, 31(4):2121--2132, 1985.

\bibitem{Foulkes2016}
W.M.C. Foulkes.
\newblock {Tight-Binding Models and Coulomb Interactions for s, p, and d
  Electrons}.
\newblock In Eva Pavarini, Erik Koch, Jeroen {Van Den Brink}, and George
  Sawatzky, editors, {\em Quantum Materials: Experiments and Theory}, volume
  Volume 6, page 420, J{\"{u}}lich, 2016. Forschungszentrum J{\"{u}}lich.

\bibitem{KhVekilov2016}
Yu~Kh Vekilov, O~M Krasilnikov, A~V Lugovskoy, and Yu~E Lozovik.
\newblock {Higher-order elastic constants and megabar pressure effects of bcc
  tungsten: Ab initio calculations}.
\newblock {\em Physical Review B}, 94(10):104114, 2016.

\bibitem{QeASA}
{\em The Atomic Spheres Approximation}, accessed December, 2019.
\newblock https://www.questaal.org/docs/code/asaoverview/.

\bibitem{Hansen1996}
N.~{Hansen} and A.~{Ostermeier}.
\newblock Adapting arbitrary normal mutation distributions in evolution
  strategies: the covariance matrix adaptation.
\newblock In {\em Proceedings of IEEE International Conference on Evolutionary
  Computation}, pages 312--317, May 1996.

\bibitem{hansen2019pycma}
Nikolaus Hansen, Youhei Akimoto, and Petr Baudis.
\newblock {CMA-ES/pycma} on {G}ithub.
\newblock Zenodo, DOI:10.5281/zenodo.2559634, February 2019.

\bibitem{Dufresne2015}
Alice Dufresne, Fabienne Ribeiro, and Guy Tr{\'{e}}glia.
\newblock {How to derive tight-binding spd potentials? Application to
  zirconium}.
\newblock {\em Journal of Physics Condensed Matter}, 27(33), aug 2015.

\bibitem{Endres2018}
Stefan~C. Endres, Carl Sandrock, and Walter~W. Focke.
\newblock {A simplicial homology algorithm for Lipschitz optimisation}.
\newblock {\em Journal of Global Optimization}, 72(2):181--217, oct 2018.

\bibitem{2019arXiv190710121V}
Pauli {Virtanen}, Ralf {Gommers}, Travis~E. {Oliphant}, Matt {Haberland}, Tyler
  {Reddy}, David {Cournapeau}, Evgeni {Burovski}, Pearu {Peterson}, Warren
  {Weckesser}, Jonathan {Bright}, St{\'e}fan~J. {van der Walt}, Matthew
  {Brett}, Joshua {Wilson}, K.~{Jarrod Millman}, Nikolay {Mayorov}, Andrew
  R.~J. {Nelson}, Eric {Jones}, Robert {Kern}, Eric {Larson}, CJ~{Carey},
  {\.I}lhan {Polat}, Yu~{Feng}, Eric~W. {Moore}, Jake {Vand erPlas}, Denis
  {Laxalde}, Josef {Perktold}, Robert {Cimrman}, Ian {Henriksen}, E.~A.
  {Quintero}, Charles~R {Harris}, Anne~M. {Archibald}, Ant{\^o}nio~H.
  {Ribeiro}, Fabian {Pedregosa}, Paul {van Mulbregt}, and SciPy 1.~0
  {Contributors}.
\newblock {SciPy 1.0--Fundamental Algorithms for Scientific Computing in
  Python}.
\newblock {\em arXiv e-prints}, page arXiv:1907.10121, Jul 2019.

\bibitem{Gao2012}
Fuchang Gao and Lixing Han.
\newblock {Implementing the Nelder-Mead simplex algorithm with adaptive
  parameters}.
\newblock {\em Computational Optimization and Applications}, 51(1):259--277,
  jan 2012.

\bibitem{Nelder1965}
J.~A. Nelder and R.~Mead.
\newblock {A Simplex Method for Function Minimization}.
\newblock {\em The Computer Journal}, 7(4):308--313, 01 1965.

\bibitem{Goodwin1989}
L~Goodwin, A~J Skinner, and D~G Pettifor.
\newblock {Generating transferable tight-binding parameters: Application to
  silicon}.
\newblock {\em EPL}, 9(7):701--706, 1989.

\bibitem{pbe1996}
John~P. Perdew, Kieron Burke, and Yue Wang.
\newblock Generalized gradient approximation for the exchange-correlation hole
  of a many-electron system.
\newblock {\em Phys. Rev. B}, 54:16533--16539, Dec 1996.

\bibitem{Brugger1964}
K.~Brugger.
\newblock {Thermodynamic definition of higher order elastic coefficients}.
\newblock {\em Physical Review}, 133(6A):6, 1964.

\bibitem{Adams2006}
J~J Adams, D~S Agosta, R~G Leisure, and H~Ledbetter.
\newblock {Elastic constants of monocrystal iron from 3 to 500 K}.
\newblock {\em Journal of Applied Physics}, 100(11):113530, 2006.

\bibitem{Rayne1961}
J.~A. Rayne and B.~S. Chandrasekhar.
\newblock {Elastic constants of iron from 4.2 to 300°K}.
\newblock {\em Physical Review}, 122(6):1714--1716, 1961.

\bibitem{Hughes1953}
D.~S. Hughes and J.~L. Kelly.
\newblock Second-order elastic deformation of solids.
\newblock {\em Phys. Rev.}, 92:1145--1149, Dec 1953.

\bibitem{choy1979elastic}
M.M. Choy, K.H. Hellwege, and A.M. Hellwege.
\newblock {\em Elastic, piezoelectric, pyroelectric, piezooptic, electrooptic
  constants, and nonlinear dielectric susceptibilities of crystals: revised and
  expanded edition of volumes III/1 and III/2}.
\newblock Numerical data and functional relationships in science and
  technology, new series: Crystal and solid state physics. Springer-Verlag,
  1979.

\bibitem{pham2010lattice}
Hieu~H Pham and Tahir Cagin.
\newblock Lattice dynamics and second and third order elastic constants of iron
  at elevated pressures.
\newblock {\em Computers, Materials \& Continua (CMC)}, 16(2):175--194, 2010.

\bibitem{Blaschke2017}
Daniel~N. Blaschke.
\newblock Averaging of elastic constants for polycrystals.
\newblock {\em Journal of Applied Physics}, 122(14):145110, 2017.

\bibitem{Alchagirov2001}
Alim~B Alchagirov, John~P Perdew, Jonathan~C Boettger, R~C Albers, and Carlos
  Fiolhais.
\newblock {Energy and pressure versus volume: Equations of state motivated by
  the stabilized jellium model}.
\newblock {\em Physical Review B - Condensed Matter and Materials Physics},
  63(22):2241151--22411516, 2001.

\bibitem{Kresse1996}
G.~Kresse and J.~Furthm{\"{u}}ller.
\newblock {Efficient iterative schemes for ab initio total-energy calculations
  using a plane-wave basis set}.
\newblock {\em Physical Review B - Condensed Matter and Materials Physics},
  54(16):11169--11186, 1996.

\bibitem{Joubert1999}
D.~Joubert.
\newblock {From ultrasoft pseudopotentials to the projector augmented-wave
  method}.
\newblock {\em Physical Review B - Condensed Matter and Materials Physics},
  59(3):1758--1775, 1999.

\bibitem{Wrobel2015}
Jan~S Wr{\'{o}}bel, Duc Nguyen-Manh, Mikhail~Yu Lavrentiev, Marek Muzyk, and
  Sergei~L Dudarev.
\newblock {Phase stability of ternary fcc and bcc Fe-Cr-Ni alloys}.
\newblock {\em Physical Review B - Condensed Matter and Materials Physics},
  91(2):24108, 2015.

\bibitem{Paxton2013}
A.~T. Paxton and C.~Els{\"{a}}sser.
\newblock {Analysis of a carbon dimer bound to a vacancy in iron using density
  functional theory and a tight binding model}.
\newblock {\em Physical Review B - Condensed Matter and Materials Physics},
  87(22), jun 2013.

\bibitem{Dragoni2018}
Daniele Dragoni, Thomas~D Daff, G{\'{a}}bor Cs{\'{a}}nyi, and Nicola Marzari.
\newblock {Achieving DFT accuracy with a machine-learning interatomic
  potential: Thermomechanics and defects in bcc ferromagnetic iron}.
\newblock {\em Physical Review Materials}, 2(1):13808, 2018.

\bibitem{Bartok2010}
Albert~P. Bart\'ok, Mike~C. Payne, Risi Kondor, and G\'abor Cs\'anyi.
\newblock Gaussian approximation potentials: The accuracy of quantum mechanics,
  without the electrons.
\newblock {\em Phys. Rev. Lett.}, 104:136403, Apr 2010.

\bibitem{Messina2013}
Luca Messina, Zhongwen Chang, and P{\"{a}}r Olsson.
\newblock {Ab initio modelling of vacancy-solute dragging in dilute irradiated
  iron-based alloys}.
\newblock {\em Nuclear Instruments and Methods in Physics Research, Section B:
  Beam Interactions with Materials and Atoms}, 303:28--32, 2013.

\bibitem{Whiting2019}
T.~M. Whiting, P.~A. Burr, D.~J.~M. King, and M.~R. Wenman.
\newblock Understanding the importance of the energetics of mn, ni, cu, si and
  vacancy triplet clusters in bcc fe.
\newblock {\em Journal of Applied Physics}, 126(11):115901, 2019.

\end{thebibliography}

\appendix
\end{document}